\documentclass[%
 aip,
 amsmath,amssymb,
 reprint,%
]{revtex4-1}

\usepackage{graphicx}
\usepackage{dcolumn}
\usepackage{bm}
\usepackage[colorlinks,linkcolor=red,anchorcolor=blue,citecolor=green,CJKbookmarks=True]{hyperref}
\usepackage[utf8]{inputenc}
\usepackage[T1]{fontenc}
\usepackage{mathptmx}
\usepackage{etoolbox}
\usepackage{multirow}
\usepackage{makecell}
\makeatletter
\def\@email#1#2{%
 \endgroup
 \patchcmd{\titleblock@produce}
  {\frontmatter@RRAPformat}
  {\frontmatter@RRAPformat{\produce@RRAP{*#1\href{mailto:#2}{#2}}}\frontmatter@RRAPformat}
  {}{}
}%
\makeatother
\begin{document}

\preprint{AIP/123-QED}

\title{Experimental study of underwater explosions below a free surface: bubble dynamics and pressure wave emission}
\author{Ming-Kang Li}
\affiliation{College of Shipbuilding Engineering, Harbin Engineering University, Harbin 150001, China;}

\author{Shiping Wang} \email{wangshiping@hrbeu.edu.cn}
\affiliation{College of Shipbuilding Engineering, Harbin Engineering University, Harbin 150001, China;}%
\author{Shuai Zhang}
\affiliation{College of Shipbuilding Engineering, Harbin Engineering University, Harbin 150001, China;}
\author{ Hemant Sagar}
\affiliation{
Institute of Ship Technology Ocean Engineering and Transport Systems, University of Duisburg-Essen, Bismarckstr. 47055, Duisburg, Germany;}

\date{\today}

\begin{abstract}
	The current work experimentally studies the complex interaction between underwater explosion (UNDEX) bubbles and a free surface. We aim to reveal the dependence of the associated physics on the key factor, namely, the dimensionless detonation depth $\gamma$ (scaled by the maximum equivalent bubble radius). Four typical bubble behavior patterns are identified with the respective range of $\gamma$: (i) bubble bursting at the free surface, (ii) bubble jetting downward, (iii) neutral collapse of the bubble, and (iv) quasi-free-field motion. {By comparison of the jet direction and the migration of the bubble centroid, a critical value of $\gamma$ is vital for ignoring the effects of the free surface on UNDEX bubbles.} Good agreements are obtained between the experimental data and the unified theory for bubble dynamics by Zhang et al.\cite{RN3019}. Additionally, the dependence of the pressure signals in the flow field on  $\gamma$ is investigated. The peak pressure, impulse, and energy dissipation in the UNDEX are investigated. 
\end{abstract}

\maketitle
\section{Introduction}\label{s:introduction}
\label{s:introduction}
Underwater explosion (UNDEX) plays a vital role in the national defense field \citep{RN2475}. However, there are many fundamental problems to be solved regarding UNDEX. Generally, the attentions in previous studies are focused on: shock wave, bubble pulse, and the evolution of the water surface. The shock wave is characterized by a high peak but a short duration of time \citep{RN2732,RN2733,RN2734}, which usually causes local damage to a floating structure. The bubble pulsation process is more complex and is highly dependent on the boundary condition\cite{WOS:L}. Contrary to the shock wave, the bubble pulse is characterized as a low-pressure magnitude but temporally for a longer duration. The impulse of the bubble pulse is thought to be at the same magnitude level as the shock wave \citep{RN2475}. When the bubble's oscillation frequency matches with the natural frequency of the marine structure, the violent resonant response may be caused \citep{RN2609,RN3015}, resulting in significant structural damage.

{ The pulsation of a bubble in an infinite fluid field can be well predicted by various analytical models \citep{RN2655,RN3014,RN2561}, such as Rayleigh-Plesset model\cite{RN2474}, Keller Miksis model\cite{RN2656} and etc. When the bubble is generated in the vortex, it assumes complex evolution patterns. Zhang et al.\citep{Chs} revealed the influence law of viscosity, surface tension and buoyancy to the vortex bubble entrainment and provided new insights into the control on vortex bubble entrainment. When the bubble is initiated near the boundary, the mutual interaction between the oscillating bubble and the boundary, the so-called Bjerknes effect, changes the bubble shape to aspherical\citep{RN20221,RN2991,RN2903,RN3017}. Usually, a strong jet\cite{RN2746} drives the bubble moving closer or farther to the boundary. Zhang et al.\cite{RN2797} originally proposed the multiple vortex ring model and discovered the mechanism of a toroidal bubble splitting near a rigid boundary. It has been confirmed by various experiments \citep{RN2827,RN2632,RN3030} and numerical simulations \citep{RN2828,RN2602,RN2634} that the Bjerknes force drives the bubble away from the boundary when the bubble is located beneath the water surface}. It was pointed out by Wang et al.\cite{RN2616} that the repulsive force is derived from a stagnation point along the symmetry axis between the top of the bubble and the free surface when the bubble contracts. According to Bernoulli's principle, a high-pressure region exists at the stagnation point which redirects the incoming flow to the downward direction. It has also been discussed in Ref.\cite{RN2607} that this jet may also have  resulted from the combined effect of less Rayleigh time at the bubble top (higher curvature at the bubble top leads to faster collapse) \citep{RN2856} and the higher universal driving force above the bubble \citep{RN2844}. The variation of the relative strength of this Bjerknes force and buoyancy renders the bubble to have different behavioral patterns. Many numerical studies have been conducted to investigate the interaction between the oscillatory bubble and free surface, including bubble shape evolution at different buoyancy parameter $\delta$ and standoff distance $\gamma$  \citep{RN2657,RN2616}, the dynamics of the two bubbles along the axis \citep{RN2602}, bubble and free surface dynamics in shallow underwater explosion \citep{RN31}, etc. 

Apart from these, some experiments have been conducted for validation \citep{RN2621,RN2614,RN2618,RN2628} as well as to investigate the phenomena that are hard for numerical simulations to clarify: bubble migration for multiple bubble oscillation cycles \citep{RN2495} and the interaction of bubble and free surface when bubble inception position is extremely close to the free surface \citep{RN2632} etc. There are generally three kinds of experimental methods to study the bubble dynamics, namely: laser-induced bubble\citep{RN3015}, spark-generated bubble\citep{RN2608}, and underwater explosion (UNDEX) bubble\citep{RN33}. The laser-induced bubble has an ideal spherical shape during growth and often these bubbles were investigated at microscale \citep{RN2737}. The spark-generated bubble \citep{RN2625} is the most widely used method as an alternative to investigating UNDEX bubble dynamics for its convenience, safety, economy, and ability to study the high-pressure bubble dynamics under some non-dimensional parameters, such as standoff parameter $\gamma$ and buoyancy parameter $\delta$ at reduced ambient pressure. However, it has been analyzed by Hung et al.\cite{RN2603} that the products of the spark-generated bubble come from the dissolved air and water, which will disintegrate into the surrounding water when the bubble collapses due to the high pressure. That results in a reduction in the energy of the bubble compared with the UNDEX bubble, especially after one bubble cycle. The pressure induced by the first shock wave is not possible to be captured for a spark-generated bubble as the detonation process is not involved. Therefore a necessity to conduct a real underwater explosion experiment remains in force. Because of the excessive cost and safety concerns, the underwater experiments for scientific investigations are usually limited to small-charge  ($R_{\rm max}\approx 0.25$ m) \citep{RN33,RN2618} or mini-charge ($R_{\rm max}\approx 0.15$ m) \citep{RN2603,RN2641}. The maximum radius of the bubble in our experiments was about 0.4 m which is larger than that in the aforementioned  literature.

There are three phenomena for a high-pressure bubble near the free surface that previously have attracted much attention: the water plume rising and splashing phenomenon, bubble dynamics patterns at different distances to the free surface, and the shock wave emission characteristics. Further two phenomena are discussed in the scope of the current study. Usually, the buoyancy parameter $\delta$ is small for the conventional experimental-scale bubble. Zhang et al.\cite{RN2495} studied bubble dynamics at variable atmospheric pressure, which showed completely different bubble dynamics at large buoyancy parameters. Brett et al.\cite{RN2631} studied the characteristics of bubble collapse pressure wave near a free surface based on a mini-charge UNDEX experiment, in which they found that the bubble pulse pressure reaches a maximum value when the bubble's migration is not observable. Apart from these, the energy dissipation mechanism is another important issue related to the bubble pulsation that was not frequently discussed and test cases only focused on the free field condition\citep{RN2736,RN2622}. There are generally three sources for energy dissipation \citep{RN2475,RN2622}, such as  heat transfer, induced turbulent flow, and the compressibility of the fluid, among which  the compressibility is thought to be the main source for an UNDEX bubble. Previous researchers usually concentrate on one aspect in their investigations. However, we are systematically investigating these issues based on large-scale UNDEX experiments in the present study.

Based on the state of the art stated above, a series of  UNDEX experiments with larger bubble radius $R_{\rm max}\approx 0.4$ m beneath the free water surface were conducted. The bubble dynamics and migration processes were captured by a high-speed camera. The temporal pressure curves of the shock wave and the bubble pulse were measured by the pressure sensors. Based on the pressure measurements, the laws of the bubble pulse peak, impulse, and the shock wave impulse with gauge distances as well as the depth of the gauge point were obtained. The energy dissipation of bubble-water systems at different standoff distances $\gamma$ was investigated by the pressure curve along with the recorded images.  { Overall, our study gives the overview of the large-scale scientific underwater explosion investigations at various relative depths. In addition, our findings regarding pressure peaks and images may be helpful for an in-depth understanding of full-scale underwater explosions and their detonation strength. }
\section{Experiment setup and data processing}
\label{S:experiment setup}
The UNDEX experiments are carried out in a $4$ $\times 4$ $\times 4\,\text{m}^3$  cubical tank made of steel wall with a thickness of about 1 cm. The cubical container has in total three windows for various purposes. There is an observation window located on one side for high-speed photography as is shown in Fig.\ref{Fig:experimental setup}(a). While the other two windows were fixed right at the neighboring walls of the tank for illumination and further observation. The main illumination we use was sunlight and the spotlights served as an auxiliary illumination. Bubble shapes were captured by the high-speed camera (Phantom V12.1) at a speed of 9150-16000 frames per second. The captured images had a resolution of 480 $\times$600 pixels with a calibrated resolution of 2 mm per pixel.  {{The lower resolution can have effects on the quality of images providing fewer details.  In order to compensate for the imaging speed, we were strict with the resolution of images. In our case, the maximum bubble size ($\sim$0.4m) was covered by 200 pixels per bubble radius which reflected overall acceptable global features of bubble dynamics.}There was a simple truss structure placed at the top of the tank. The explosive charge was attached to a string with its upper end fixed on the center of the truss and its lower end attached with a counter weight to keep the string straight. The captured images containing the shape of the bubble are processed further to obtain clarity about the bubble dynamics. The charge type used in this study is RDX (Research Department Explosive), which is detonated by an electric detonator with consideration of safety. { Two piezoelectric pressure sensors ($\text{PCB}^{\textcircled{c}}$) were used to measure the transient pressure. Their resolution  is 0.07 kPa which is significantly small compared with the shock wave and the bubble pulse, as will be shown in Fig.\ref{Fig:free field pressure} and Fig.\ref{Fig:free field pressure2}.}

After the explosion, a bubble filled with high-pressure explosion gases was formed, during which the chemical energy of the charge turns into the potential energy inside the bubble. At the effect of balance between internal bubble pressure and external pressure at the bubble wall, the bubble would experience several cycles of expansion and contraction till energy is entirely dissipated. The bubble keeps its spherical shape during the expansion phase after detonation. The bubble would enter the stage of collapse when it contracts to a small volume. A high-speed jet is formed at the end of collapse and rebounds. During this process, the bubble does not remain spherical anymore. Hence the equivalent radius $R$ is used in this study, which is obtained by estimating the volume of the bubble $V$ and then with the formula $R=\sqrt[3]{3V/4\pi}$. The volume of the bubble is obtained in a slice-by-slice manner, see Fig.\ref{Fig:experimental setup}(b). For volume estimation, we assumed the bubble shape to be axisymmetric during the first two bubble oscillations. Accordingly, we establish a coordinate system having the origin at the charge center and sliced the bubble image from top to bottom into about 20 to 30 sections. For each section, the shape is assumed to be the frustum of a cone. Hence the volume of the bubble is obtained by 
\begin{equation}
\label{eq:volume}
V=\sum_{1}^{n}\pi (z_{i}-z_{i+1})(r_{i}^2+r_{i}r_{i+1}+r_{i+1}^2)/3
\end{equation}
where $r$ and $z$ are radial and vertical coordinates respectively. As the migration of the bubble will be considered in the later section, the centroid of the bubble is calculated by

\begin{equation}
\label{eq:centroid}
Z=
\frac{\begin{aligned}\sum_{1}^{n}[\pi (z_{i}-z_{i+1})^2(3r_{i}^2+2r_{i}r_{i+1}+r_{i+1}^2)/12\\
	+\pi z_{i+1} (z_{i}-z_{i+1})(r_{i}^2+r_{i}r_{i+1}+r_{i+1}^2)/3]
\end{aligned}
}{V}
\end{equation}

\begin{figure*}
	\centering\includegraphics[width=16cm]
	{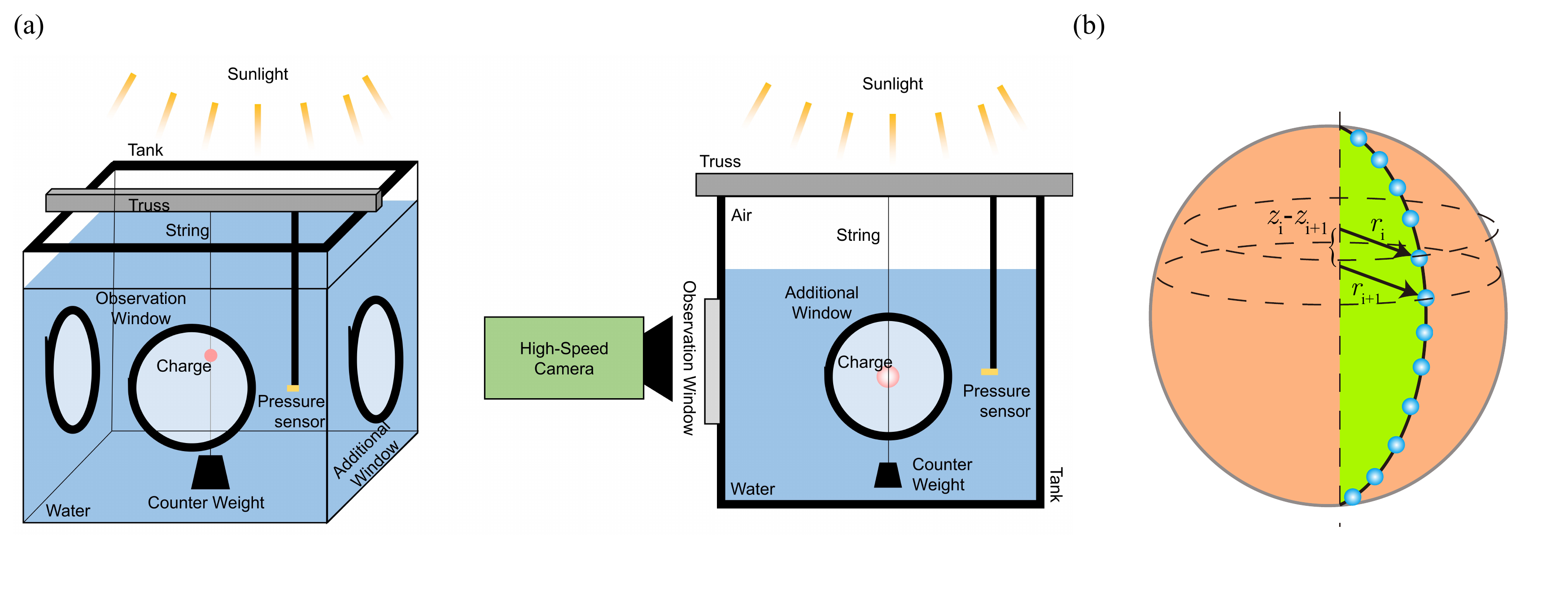}\\\caption{(a) A sketch for the experimental setup. A truss structure is placed at the top of the tank. A string hangs straight with its upper end fixed at the center of the truss structure and the lower end attached with a counter weight. The explosive charge is attached to the string at the presetting depth. (b) The sketch for calculating the volume and centroid of the bubble in a slice-by-slice manner. A local coordinate is established at the detonation point. The bubble image is divided into 20 to 30 parts and calculate their volume and their moment to the coordinate axis respectively.}\label{Fig:experimental setup}	
\end{figure*}

The movement of a high-pressure bubble near the free surface is influenced by two forces: buoyancy and Bjerknes force. These two forces are in opposite directions and their relative strength affects the bubble dynamics near the free surface. As Bjerknes force is strongly influenced by the bubble distance to the free surface, a standoff parameter $\gamma$ is introduced as follows:
\begin{equation}
\label{eq:standoff}
\gamma=\frac{H}{R_{\rm{max}}},
\end{equation}
where $H$ denotes the detonation depth and $R_{\rm{max}}$ is the maximum equivalent radius of the bubble at the respective water depth. { The surface tension effect is an important quantity that affects the bubble dyanmics\cite{WOS:000521052700001,WOS:000807121500001}. To take the surface tension into account, the Weber number is introduced\cite{RN3037}: $We=R_{\text{max}} P_{\infty}/\sigma$, in which $\sigma$ is the surface tension coefficient. Then if we take $R_{\text{max}}=0.4$ m,  $P_{\infty}=1\times 10^5$ Pa and $\sigma=7.28\times 10^{-2}$ N/m into above equation, then we get $We=5.5\times 10^5$. In Li et al.\cite{RN3037}(2016), the authors showed that when $We=1.2\times10^3$, the volume evolution is identical to that at  $We=\infty$. Thus in our UNDEX experiments, the surface tension effect can be ignored for the significantly higher Weber number.} To measure the strength of buoyancy, a buoyancy parameter \citep{RN2495} is defined as :
\begin{equation}
\label{eq:buoyancy parameter}
\delta=\sqrt{\frac{\rho g R_{\rm{max}}}{p_{\infty}}} ,
\end{equation}
where $\rho$ and $g$ denote the density of water and gravity respectively. And $p_{\infty}$ is the ambient pressure consisting of the atmospheric pressure and still water pressure at the initial charge center. Throughout this paper, the non-dimensional length and time, which are denoted by the superscript ``*'', are scaled by $R_{\rm{max}}$ and $R_{\rm{max}}(\rho/\Delta p)^{1/2}$, respectively.
\section{Theoretical model}
\label{s:theory}
Zhang et al.\cite{RN3019} for the first time proposed a unified equation for bubble dynamics which simultaneously considers boundaries, bubble interactions, ambient flow field, gravity, bubble migration, fluid compressibility, etc. Zhang's equation\cite{RN3019} is a significant breakthrough and an epoch-making milestone in the field of theoretical research of bubble dynamics after Rayleigh-Plesset equation\cite{RN2474,RN3043} (1917, 1949), Gilmore equation\cite{RN3044} (1952) and Keller equation\cite{RN2656,Keller1980} (1956, 1980). The oscillation and migration of the bubble in a compressible fluid field can be described by their unified bubble equation in an elegant mathematical form as:
\begin{equation}
\label{eq:unified}
\left( \frac{C-\dot{R}}{R}+\frac{\text{d}}{\text{d}t}  \right) \left[\frac{R^2}{C}\left( \frac{1}{2}\dot{R}^2+\frac{1}{4}v^2+h  \right)    \right]=2R\dot{R}^2+R^2\ddot{R}\, ,
\end{equation}
where each dot on the variable denotes taking the time derivative one time. $R$, $C$, $v$, and $h$ are bubble radius, sound speed, migration velocity, and the enthalpy difference at the bubble surface respectively. { The above equation is coupled with the following bubble migration equation:
\begin{equation}
\label{eq:migration}
C_aR\dot{v}+(3C_a\dot{R}+\dot{C}_{a}R)v-gR+\frac{3}{8}C_dS(v)=0
\end{equation}
where $C_a$ is the added mass coefficient, $C_d$ is the drag coefficient, $g$ is the gravity, and $S(\cdot)=(\cdot )|\cdot|$ is the signed square operator. }In the unified equation for the underwater explosion bubble, the shock wave propagation is considered. The initial conditions for the bubble expansion are obtained by solving the Euler equations, for more details, see  Zhang et al.\cite{RN3019}.

As is shown in Eq.\eqref{eq:migration}, when further solving the bubble oscillation equation, the drag coefficient $C_d$ and the added mass coefficient $C_a$ need to be determined. Here a preliminary experiment was conducted to justify the reliability of the current experiments as well as to determine the proper values for $C_a$ and $C_d$. A free field experiment was conducted with 10 grams (about 13 grams of equivalent TNT) of explosive charge at 2 m depth. Two pressure sensors were located at a radial distance of 1.11 m and 1.75 m respectively from the charge. The time series of representative instances of the bubble dynamics are shown in Fig.\ref{Fig:free field}. 

During the first cycle of the bubble (frames 1-7), the shock wave reached the surface of the observation window right after the detonation, which caused cavitation on the window surface (see frame 2 of Fig.\ref{Fig:free field}). The bubble expanded to its maximum volume at frame 3, and then it got into the contraction phase. On the edge of collapse, the bubble shape had become aspherical (see frame 6 of Fig.\ref{Fig:free field}). The bottom part of the bubble contracted quicker than the upper part, thereupon, it got flat first. Thereafter, an upward jet was formed at the bottom and then it threaded through the bubble. Finally, the jet penetrated the upper side of the bubble wall nearly at the same time when the bubble contracted to the minimum volume. The bubble split into two parts: the upper bubble bulk and the lower toroidal bubble (see the enlarged view in frame 8 of Fig.\ref{Fig:free field}). After that two distinct bubbles started to rebound, and a pulsation pressure wave was generated by the bubbles. When the pressure wave  reached the surface of the window, the cavitation was caused on the window again (see frame 9). Simultaneously, two bubbles were observed to coalesce. A protrusion was formed at the upper end which was due to the violent upward jet. The jet carried the mixture of bubble and water and penetrated the upper end of the bubble wall. At the end of  the second cycle (frames 14-16), there was no obvious jet observed as it did in the first cycle because of the opacity of the bubble cloud. The bottom surface of the bubble got flat and collapsed faster in the second cycle, which indicated that buoyancy still influenced the bubble dynamics.
\begin{figure*}
	\centering\includegraphics[width=14cm]
	{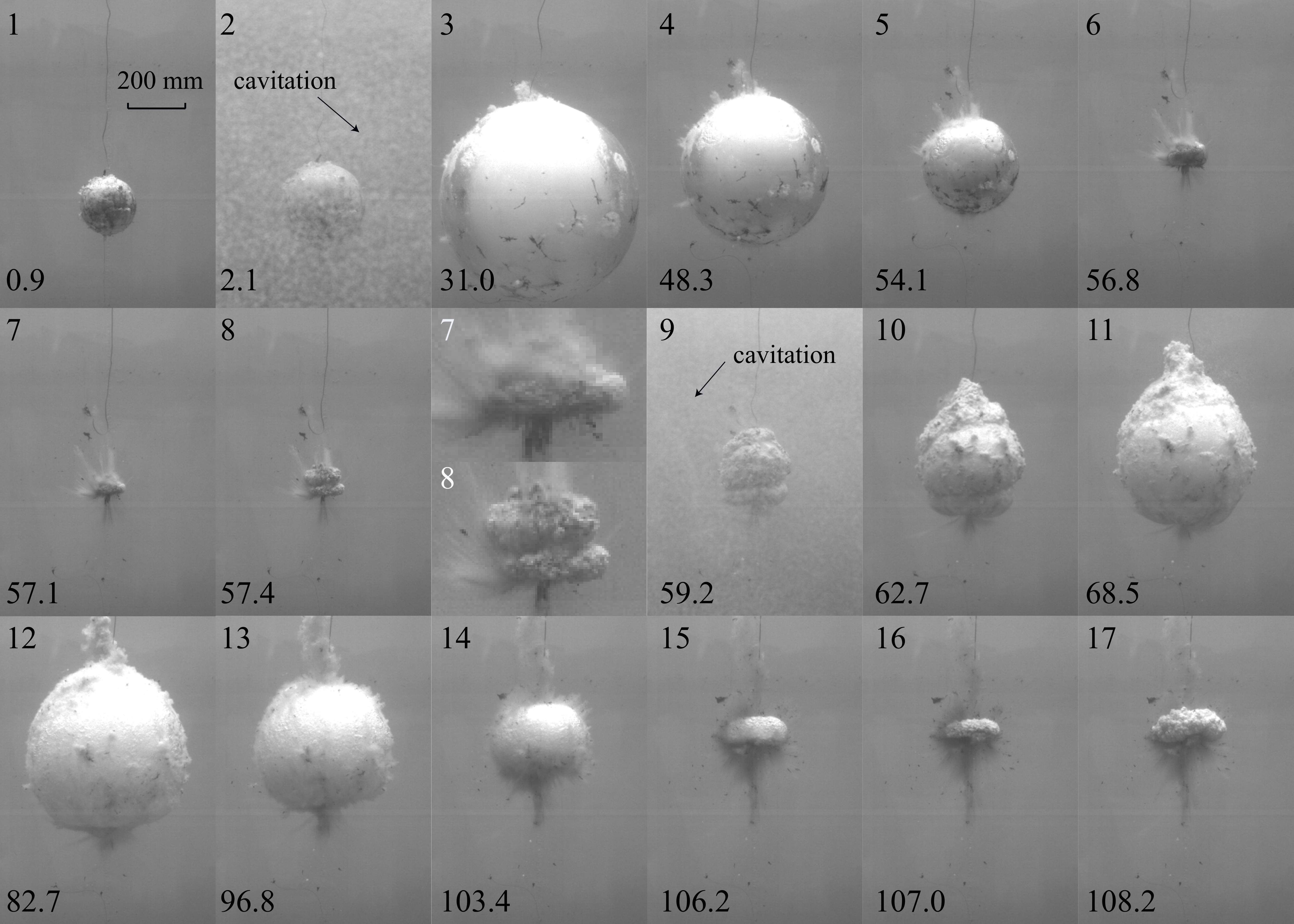}\\\caption{The bubble dynamics within two bubble's oscillation in the free field at a depth of 2 m. The black frame number is marked on the northwest corner and the corresponding physical time with unit ms is marked on the southwest corner on each frame hereafter if not specifically illustrated. The white frame number denotes the enlarged view of the bubble's shape at the respectively time. At the transformation of expansion and contraction, the bubble moves upwards at the effect of buoyancy. At the minimum volume (frame 8 and 16), a pressure wave is generated which causes cavitation on the window.}\label{Fig:free field} 
\end{figure*}

The maximum radius in the free field experiment is 35.0 cm, which is small compared to the distance from the detonation center to the boundary (2 m). The time history of the equivalent radius of the bubble and the pressure measured by pressure sensors at two probes are shown in Fig.\ref{Fig:free field period}, Fig.\ref{Fig:free field pressure} and Fig.\ref{Fig:free field pressure2}, respectively. The setups of these two pressure sensors are shown in Table.\ref{table:pressure measurement}. In the equivalent radius time history curve, the experimental data and the solution from the unified equation for UNDEX bubble\cite{RN3019} ($C_d=1.5$, $C_a=0.2$) are compared. In terms of radius, the two curves match well in the aspect of the maximum radius and the period in the first cycle. During the second period, some discrepancies present with smaller bubble sizes and shorter oscillation periods in experiments. In terms of pressure peaks and trends, the theoretical model matches well with the experimental data. The comparisons show that the data of this simple experiment can be reliably validated by the theoretical model. Therefore one can rely on the experimental data obtained afterward based on the same experimental setups. 
\begin{figure}
	\centering\includegraphics[width=8cm]
	{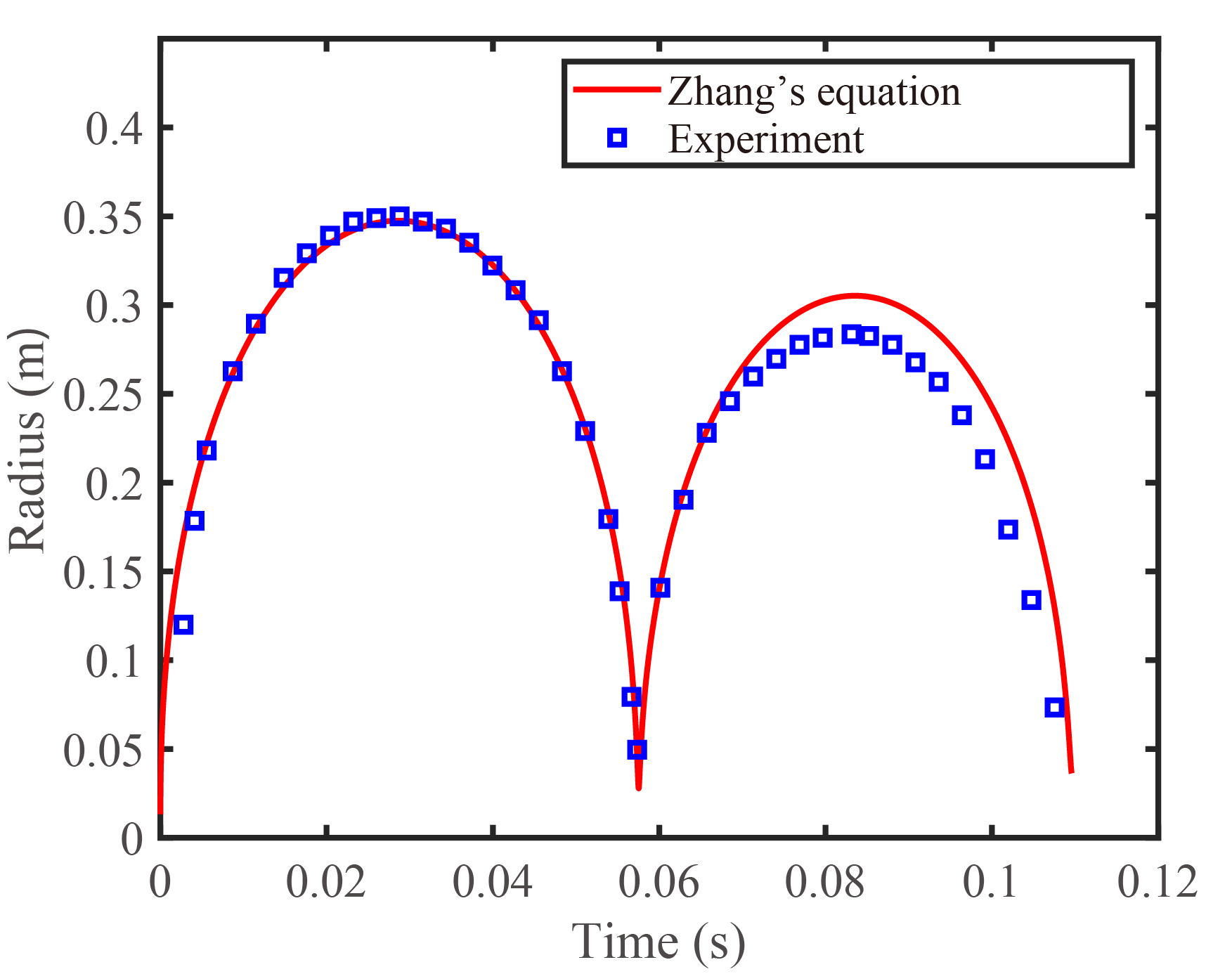}\\\caption{The time variation of the equivalent radius of the bubble from the experiment and Zhang equation\cite{RN3019} for underwater explosion bubble with $C_d=1.5$, $C_a=0.2$.}\label{Fig:free field period}	
\end{figure}
\begin{figure*}
	\centering\includegraphics[width=15cm]
	{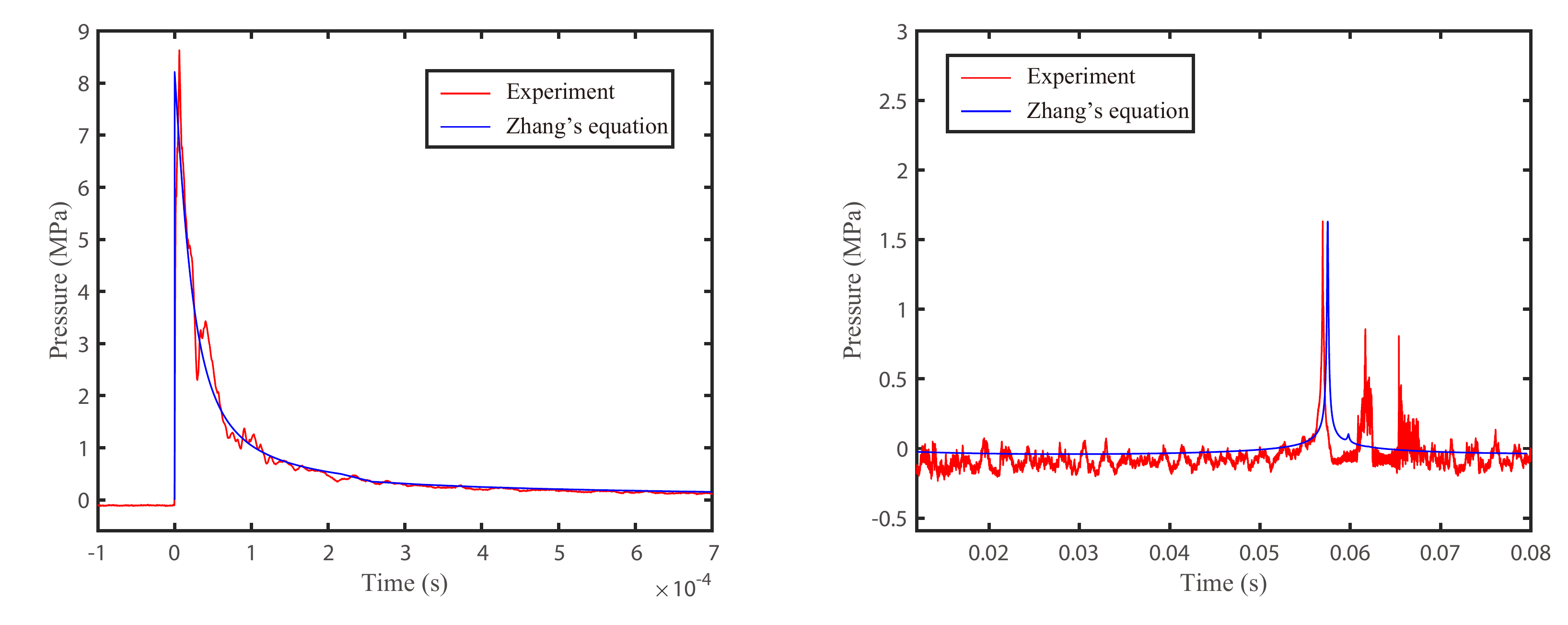}\\\caption{The comparison of the recorded pressure curves and the results from Zhang equation\cite{RN3019} with $C_d=1.5$, $C_a=0.2$ at the position $r=1.11$ m. Left: shock wave stage Right: bubble oscillation stage.}\label{Fig:free field pressure}	
\end{figure*}
\begin{figure*}
	\centering\includegraphics[width=15cm]
	{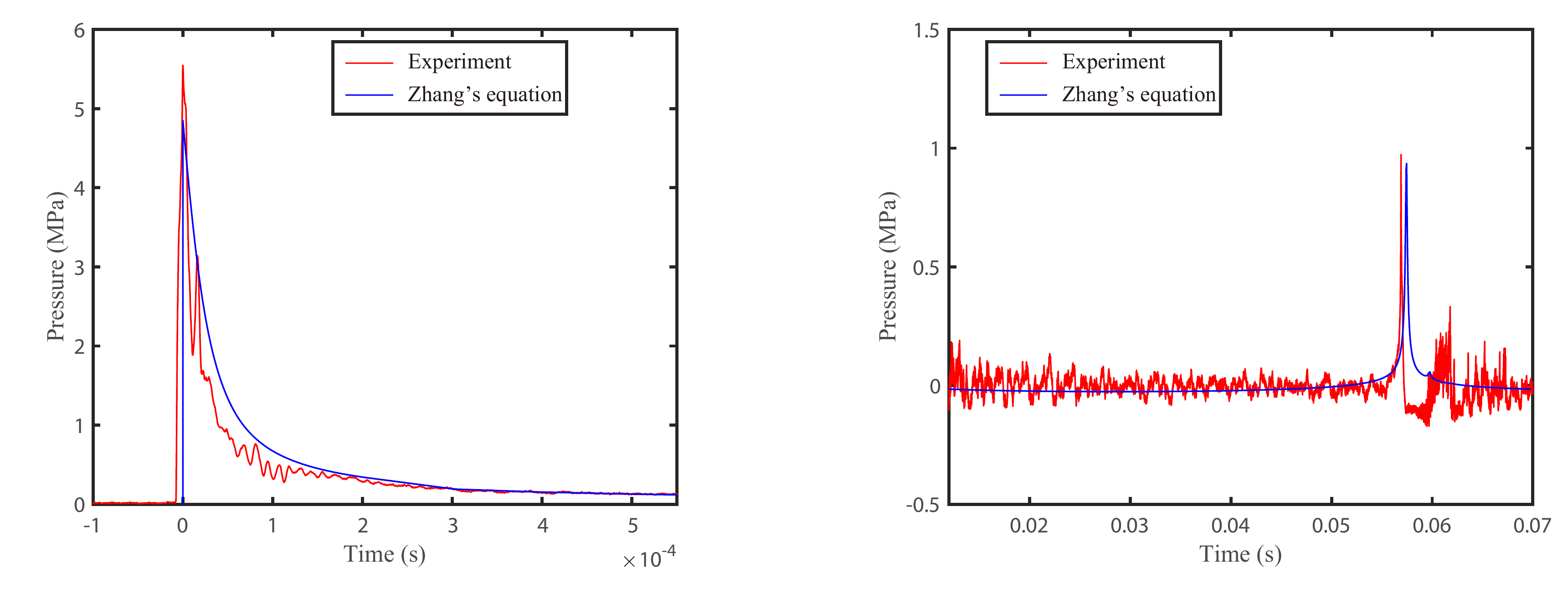}\\\caption{The comparison of the recorded pressure curves and the results from Zhang equation\cite{RN3019} with $C_d=1.5$, $C_a=0.2$ at the position $r=1.75$ m. Left: shock wave stage Right: bubble oscillation stage.}\label{Fig:free field pressure2}	
\end{figure*}

{
To further discuss the energy loss issues of bubble collapse in section\ref{s:energy loss}, the illustration of the energy of the bubble is briefed here. The process of detonation and the transformation of chemical energy to the internal energy of explosive products are not considered in the present study. According to the conservation of energy, the total energy of the bubble-water system can be written as: 
\begin{equation}
E=E_k+E_p
\end{equation}
in which $E_k$ and $E_p$ are kinetic energy and potential energy, respectively. According to Li et al.\cite{RN2618}(2019) and Tian et al.\cite{RN2974} (2021), the potential energy can be further extended in the non-dimensional form as:
\begin{equation}
\label{eq:energy}
E_p=V(1-\delta^2 z_c)+\frac{\varepsilon V}{\gamma-1}\left(\frac{V_0}{V}\right)^{\gamma}
\end{equation}
in which $z_c$, $\varepsilon$, $\gamma$, $V_0$, and $V$ are the vertical locations of the bubble center, the strength parameter, the specific heat ratio, the equivalent initial volume of the bubble and the transient volume of bubble, respectively. The first term on the right-hand side of Eq.\eqref{eq:energy} denotes the gravitational potential $E_p^g$ of water and the second term is the internal energy  $E_p^b$ of the bubble. As the mass of a bubble is significantly small, its kinetic energy and gravitational potential are usually not taken into consideration. 

The oscillation of the bubble is accompanied by the inter-transformation of the energies mentioned above as well as the energy loss taken by the pressure wave. For ease of analysis, we assume that the energy of the bubble-water system is conserved during each cycle, and energy is only lost at the beginning of the rebounding phase. At the start, the internal energy $E_p^b$ of explosive gas reaches its maximum. By doing work to the external water, $E_p^b$ is transformed to $E_k$ and $E_p^g$.  Because of the inertia of bubble expansion, the bubble will continue to expand even when the internal pressure equals the ambient pressure. During this process, $E_k$ and $E_p^b$ all transform to $E_p^g$. When the bubble expands to its maximum volume, $E_k$ is assumed to be zero, which makes the potential energy take up the majority. To analyze the magnitude of gravitational potential, the experimental parameters are taken into Eq.\eqref{eq:energy}. As $\delta^2 \approx 0.03$, hence $E_p^g \approx V$. The volume ratio $V_0/V$ in the second term of Eq.\eqref{eq:energy} can be referred to the theoretical model, see Fig.\ref{Fig:free field period}. As $R_0/R_{\rm max} \approx 0.1$, $V_0/V \approx 0.001$, which indicates that $E \approx V$ when the bubble expands to its maximum volume. The overall energy of the system in each cycle can be measured by the maximum volume of the bubble respectively, which is referred to as the volume-based approach in this paper. It has also been pointed out by Lee et al.\cite{RN2622} (2007) that the cube of the bubble oscillation period $T^3$ can also indicate the energy of the bubble, which is referred to as the period-based approach. These two approaches are principally similar if the spherical bubble model is assumed. 

To quantitatively investigate the portion of the energy carried by the pressure wave, the recorded pressure curve is utilized via the following formula \citep{RN2475}:
\begin{equation}
\label{eq:cal energy}
E_w=\frac{4\pi r^2}{\rho c}\int P^2 dt,
\end{equation} where $r$ is the distance from the gauge point to the detonation position and $P$ is the excess pressure. Eq.\eqref{eq:cal energy} will be used in the discussion about energy loss issues in section\ref{s:energy loss}}.

\section{Bubble dynamics patterns near free surface}
\label{s:results}
This section elaborates on a series of experiments close to the water surface conducted to investigate the dynamics of the UNDEX bubble, in which 10 grams of RDX charge is used for all the experimental cases. Twelve different standoff distances are investigated including one repeated experiment at $H=0.3$ m and two repeated experiments at $H=0.8$ m, detailed information on the experimental test cases can be found in Table.\ref{table:experiment case}. Four bubble dynamics patterns are observed in our experiments, which are described and illustrated below.
\begin{table}
	\centering
	\caption{Experiment cases and parameters}
	\begin{ruledtabular}
	\begin{tabular}{cccc}	
		NO.&Depth $H$  (m) &Maximum radius $R_{\rm{max}}$  (m) &Standoff parameter $\gamma$\\
		\hline
		1& 0.0&0.39&0.0\\
		2& 0.05&0.37&0.14\\
		3& 0.15&0.38&0.41\\
		4& 0.25&0.38&0.66\\
		5& 0.3&0.4&0.75\\
		6& 0.3&0.4&0.75\\
		7&0.5&0.39& 1.28\\
		8& 0.7&0.39&1.79\\
		9&0.8&0.39& 2.05\\
		10&0.8&0.38& 2.11\\
		11&0.8&0.38& 2.11\\
		12&0.9&0.39& 2.31\\
		13&1.0&0.38& 2.63\\
		14&1.2&0.37& 3.16\\
		15&1.3&0.37&3.47\\
	\end{tabular}
	\end{ruledtabular}
	\label{table:experiment case}
\end{table}

\subsection{Bubble bursting at free surface}
\label{ss:burst}
We observed the bubble bursting phenomenon at the free surface when the detonation depth is significantly small ($\gamma<0.41$). The bubble pulsation was not observed and the bubble assumed a half-sphere in the first oscillation cycle. Hence, the $R_{\rm max}$ in Table.\ref{table:experiment case} is half of the maximum horizontal width $R_{\rm w}$ during the entire process. Fig. \ref{Fig:free surface0} presents the bubble dynamics for $\gamma=0.14$. Instantly after the detonation, a significant amount of bulk cavitation could be observed in the water (frame 1 of Fig. \ref{Fig:free surface0}). The bubble boundary appeared to be crystal clear at the early stage of the expansion phase (frames 2-3, Fig. \ref{Fig:free surface0}). The mixture of explosive gas and vapor inside the bubble was sprayed out, which is an important characteristic of a UNDEX bubble bursting at the surface. Later on, the bubble wall turned opaque in frame 4. We think that it may be caused by the impact of the falling water particle inside the bubble's opened cavity and their splashing on the bubble's wall. We found that as $\gamma$ increases, the time required for the formation of opacity is prolonged. The non-dimensional time at which the opacity takes place at the bubble wall is 0.22, 0.57, 0.92, 1.12 for $\gamma$=0.14, 0.41, 0.66, and 0.75, respectively. This phenomenon is not typical of the bubble bursting as it also occurred at $\gamma=0.66$ and 0.75 during the bubble's contraction stage, in which the bubble remained intact. We have conducted the repeated experiment at $\gamma=0.75$ where we did not find the presence of this opacity. The uncertainty in observing the opacity may conclude that $\gamma=0.75$ approaches the critical value for the formation of such opacity. The formation of opacity may affect the characteristics of the bubble's pulsation pressure significantly but doesn't seem to influence the migration of the bubble centroid, which will be shown in Fig.\ref{Fig:migration} and Fig.\ref{Fig:com0-3} and elaborated later.
\begin{figure*}
	\centering\includegraphics[width=12cm]
	{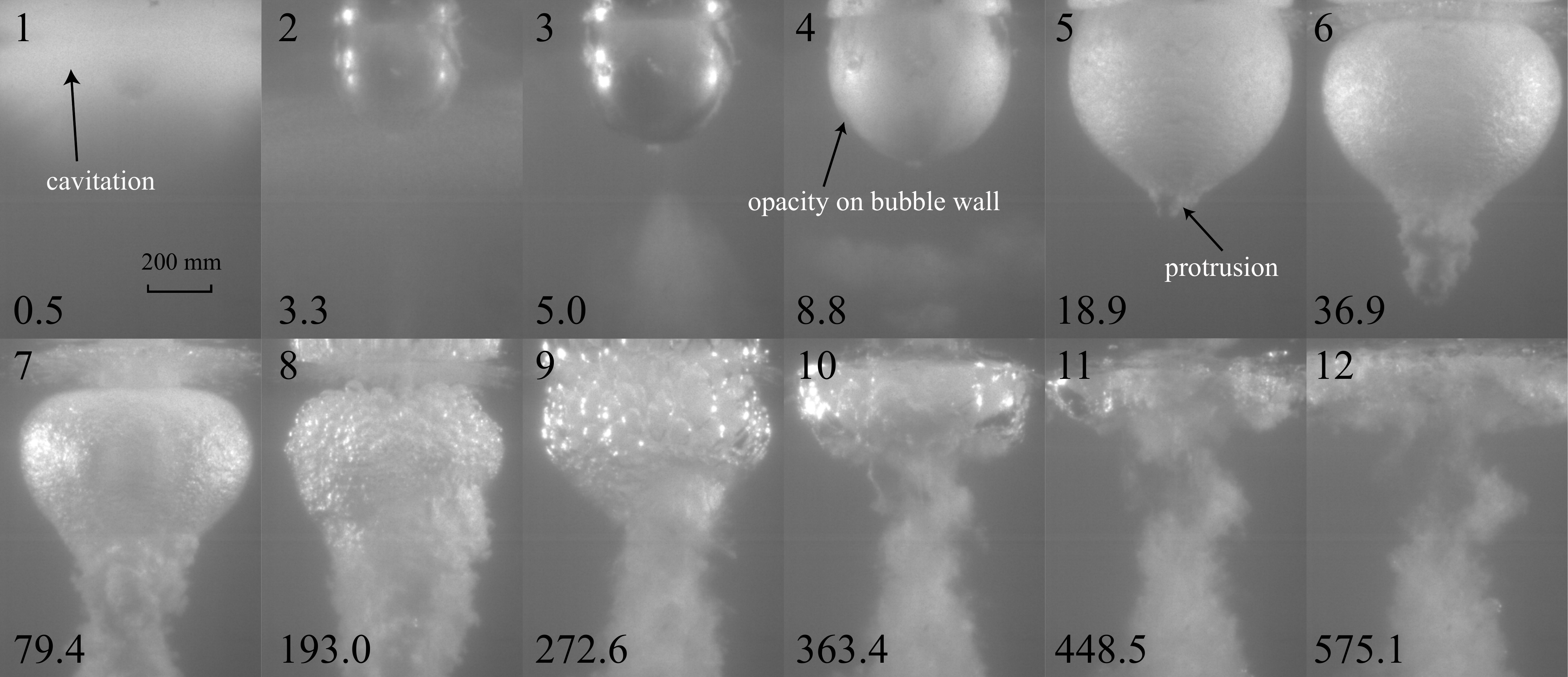}\\\caption{The bubble dynamics of the bubble bursting at the standoff parameter $\gamma$=0.14. The cavitation took place at the water surface instantly after the detonation (frame 1). Soon the bubble jetted downward (frames 5-6) and became a cloud of water and air (frame 7). There was no obvious pulsation phenomenon of the bubble in the bubble bursting pattern.}\label{Fig:free surface0}	
\end{figure*}
After the water droplets splashed over the bubble's bottom wall, a protrusion was formed at the bottom (frame 5 of Fig.\ref{Fig:free surface0}) and penetrated the bubble wall. It has been analyzed by Tian et al.(2018)\cite{RN2405} that this phenomenon occurs due to the breaking and re-closure of the bubble. After the bubble breaks, the airflow can make the displaced water at the surface join together along the vertical axis. The rejoined water impacts each other and two violent opposite jets are formed simultaneously at the surface. The upper jet becomes the so-called ``water spike'' and the lower part results in the bubble wall penetration as well as the formation of opacity on the bubble wall. This phenomenon is sensitive to the detonation depth when $\gamma\approx 0$. According to our observation, if the charge is located with its upper end touching the water surface, this protrusion can be observed. If the charge is located with its lower end touching the water surface, this protrusion is not formed. 

As the protrusion developed further and moved downward, the whole bubble became a cloud of bubbles (frames 7-9, Fig.\ref{Fig:free surface0}) with no distinct continuous boundary. The upper part of the bubble moved upward towards the free surface slowly and the detached bottom part moved more quickly, see frames 10-12 of Fig.\ref{Fig:free surface0}. During this process, the upper part of the bubble continued to rise. The gas inside the upper part of the bubble ultimately leaked into the atmosphere. The standoff distance was so small that the water layer above the bubble didn't have the potential to suppress the expansion of the high-pressure gas inside the bubble thereby breaking out of the top surface. On the one hand, high-temperature explosive gas leaked into the atmosphere; consecutively, the surrounding air at atmospheric pressure flowed into the bubble from opened part of the bubble. The mechanism from these two aspects ultimately reduces the total energy content of the bubble significantly. According to our pressure measurements, there was no pulsed pressure recorded during the bubble bursting by the sensor. 

The time history of variation of the bubble width at different $\gamma$ is shown in Fig.\ref{Fig:width} and compared with the solution from the Rayleigh-Plesset equation (RP equation) and Zhang equation\cite{RN3019}. It is found that the experimental data match well with the RP equation in the early stage ($t*<0.1$). After that, the data of all three experiments deviate from analytical solutions. As $\gamma$ increases, the deviation in time increases. After the stability point, the speed for widening in experiments is higher than the RP equation. To account for this phenomenon, we postulate that the leakage of gases does not take place instantly. It needs some time for the explosive gas to get out to the external atmosphere. Therefore, at the earlier stage of bubble expansion, there are no significant differences between the experimental data and the analytical estimations. The internal pressure of the bubble quickly falls below the ambient pressure for an intact bubble and the expansion speed decelerates. While for the bursting bubble, due to the interaction with the atmosphere, the air flows into the bubble which increases the internal pressure of the bubble. Consequently, the time for the internal pressure to be lower than the ambient pressure is supposed to be delayed which explains the faster expansion speed of the bursting bubble. As the free surface effect has been considered in Zhang equation, its expansion time history is closer to the experimental data, which shows the improvement of their proposed equation.

\begin{figure}
	\centering\includegraphics[width=7cm]
	{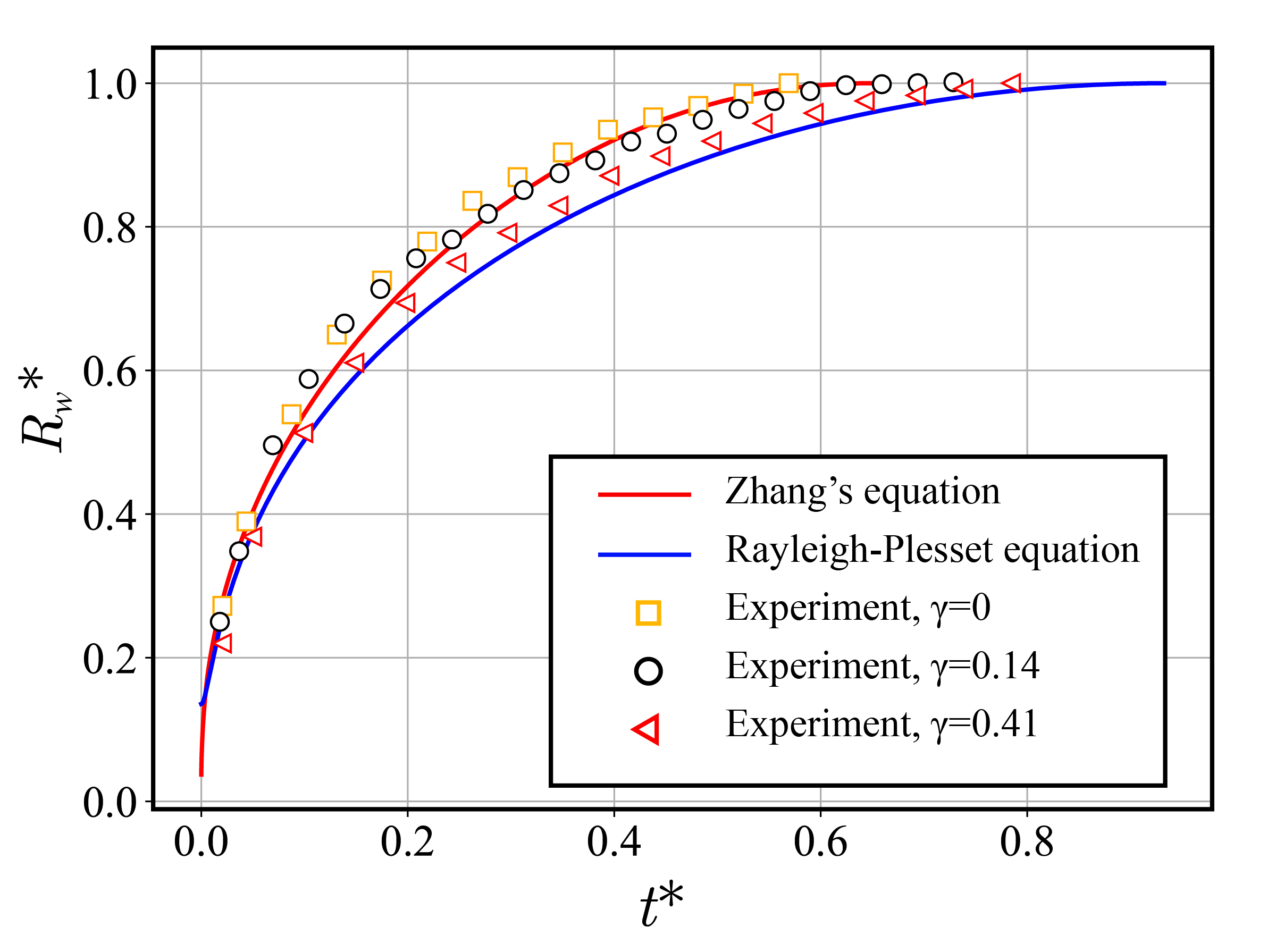}\\\caption{The growth of bubble width $R_{\rm w}$ scaled by $R_{\rm max}$ for a bursting bubble at the free surface. The analytical solutions from Rayleigh-Plesset equation\citep{RN2474} and Zhang equation\cite{RN3019} are added for comparison.}\label{Fig:width}	
\end{figure}
\subsection{Jetting downward}
When the detonation depth is higher, the sphericity and integrity of the bubble were maintained during the first oscillation cycle. From now on, the bubble assumed the pulsation characteristic. Fig.\ref{Fig:free surface1.6} shows the captured sequences of the bubble shape evolution at $\gamma=1.79$. After the bubble expanded to its maximum volume (frame 3, fig.\ref{Fig:free surface1.6}), it started to contract, during which the upper surface of the bubble got flat first (frame 4, fig.\ref{Fig:free surface1.6}). Later on, a downward re-entrant jet was then supposed to be formed (frames 5-6 ) with the help of massive bubble experiments of other sources and numerical simulations. When the bubble collapsed to its minimum volume, the jet penetrated the bubble wall and the bubble became a toroidal bubble (frames 7-8, fig.\ref{Fig:free surface1.6}). The jet carried some portion of the gas along with it and a small bubble cloud was thereupon separated from the bulk bubble. This small bubble attained high-frequency pulsation characteristics than the bulk bubble. After four cycles of pulsation, its pulsation characteristics had been significantly weakened compared with the bulk bubble. This indicates that the kinetic energy of the small bubble has nearly vanished. The bubble separation phenomenon was observed at $\gamma=1.79$, where the main bubble was divided into an upper bulk bubble and a lower detached bubble. It has influenced the energy dissipation up to a certain extent, which will be described in the section\ref{s:energy loss}. At the rebounding stage, the toroidal bubble coalesced into one single-connected bubble (frame 9). At the end of the second oscillation cycle, a downward jet was formed resulting in the migration of the bubble constantly away from the free surface. At this stage, the bubble’s dynamics  were mainly influenced by the Bjerknes force resulting from the surface, so that the bubble was constantly repelled away from the free surface downwards. This dynamics pattern is characterized as constant downward migration of bubble in our experiments, which covers the test cases of standoff parameter from $\gamma$= 0.66 to $\gamma$= 1.79.

\begin{figure*}
	\centering\includegraphics[width=12cm]
	{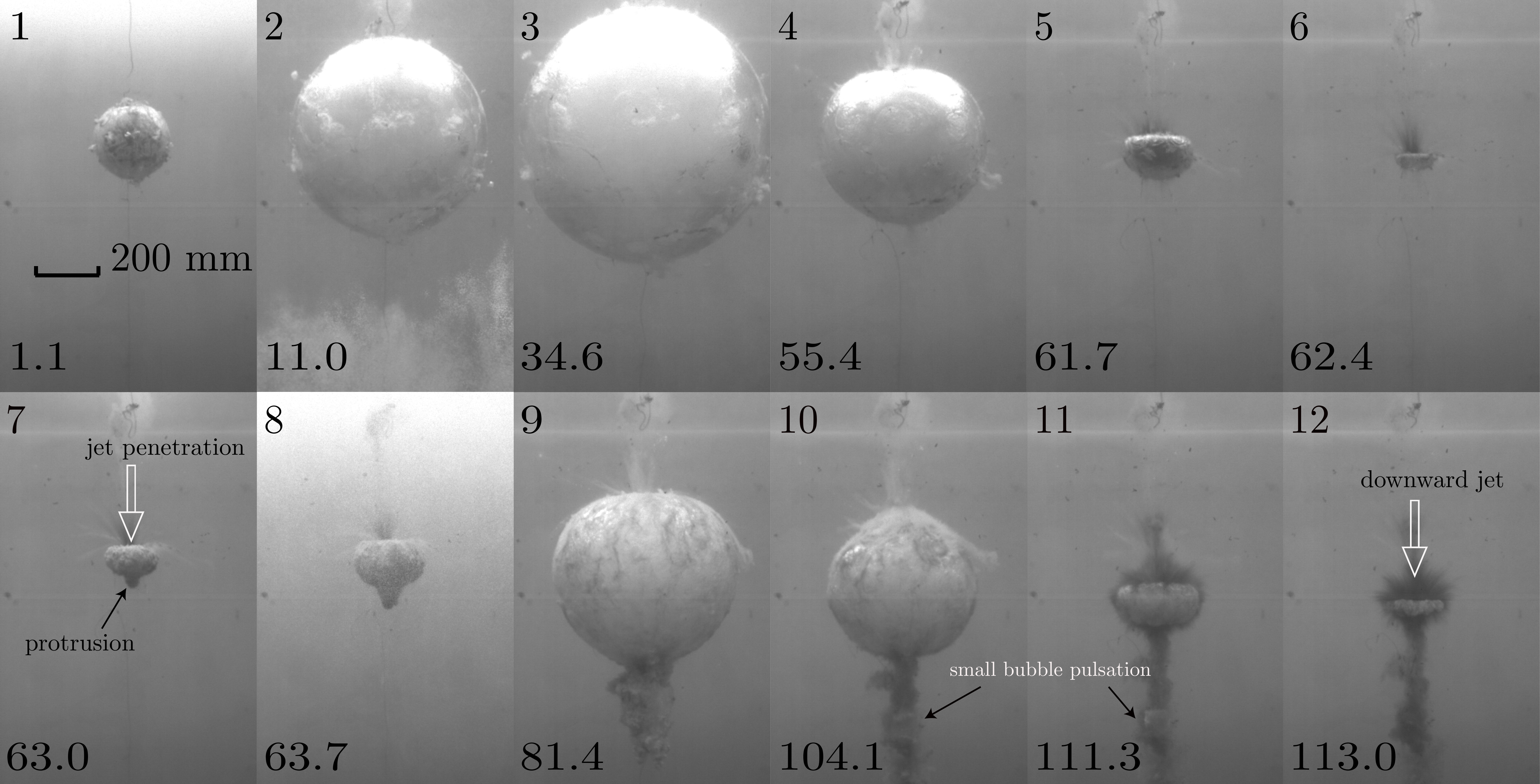}\\\caption{The bubble dynamics at the standoff parameter $\gamma$=1.79.  When the bubble reached to its minimum volume (frame 7), a downward jet penetrated the bubble wall, and carried some portion of bubble content with it (frames 7-9). The carried small bubble oscillates much faster than the bubble bulk (frames 10-11).}\label{Fig:free surface1.6}	
\end{figure*}

\subsection{Neutral collapse}
We increased the detonation depth further so that the balance between Bjerknes force and the buoyancy can be obtained at $H = 0.8$m ($\gamma\approx 2.1$). In such condition, the dynamics of the bubble is expected to be more complex. The experiments of this case have been repeated three times and the maximum equivalent bubble radius, the bubble oscillation period and the time history of the bubble’s equivalent radius are remarkably similar. Some discrepancies exist among the three repeated experiments at the collapse stage where the bubble attains its minimum volume at which instabilities are significant.

Fig.\ref{Fig:free surface1.8-1} shows bubble evolution which will be referred to as type I in the scope of the study. It can be seen that at the end of the first cycle (frames 6-8, fig. \ref{Fig:free surface1.8-1}), the  bubble shrank horizontally faster which resulted in an annular jet. After jetting the bubble split into two separate parts. These two distinct parts started to coalesce along the line that two bubbles contact (frames 9-10) into a single bulk cavity at the rebounding stage. At the end of the second oscillation cycle, the bubble attained a flat lateral oval shape (see frame 13). In the third oscillation period, the annular jet was again formed (frames 17, 18). By the time the bubble split, two opposite jets are formed pushing two individual parts away from each other (see frame 18). Fig.\ref{Fig:free surface1.8-2} shows another type of neutral collapse pattern referred to as type II. It shows that the bubble nearly collapses spherically at the end of first oscillation cycle, and no significant annular jetting phenomenon was observed during the entire collapse process. For the third variant (type III), a weak downward jet was observed every time, when the bubble collapsed and the bubble migrated slightly downward. The comparison among the time history of the equivalent radius of the bubble for these three types of bubble dynamics is presented in Fig.\ref{Fig:r neutral collapse}. All three curves of equivalent bubble radius are nearly identical except for a few discrepancies in the second cycle. The difference observed in the maximum equivalent radius in the second cycle postulates that some energy loss occurred during the second oscillation cycle, which will be discussed in the section\ref{s:energy loss}. Though the bubble dynamics patterns observed in images of Fig.\ref{Fig:free surface1.8-1} and Fig.\ref{Fig:free surface1.8-2} at H = 0.8 m are a bit different, the variation in equivalent radius over time is nearly the same. It indicates that bubble dynamics are unstable when the bubble collapses to the minimum volume around this point. The source of this instability might be due to an imbalance between Bjerknes force and buoyancy force in terms of magnitude. Small perturbations in the experimental setup might have also caused the bubble to evolve differently.

\begin{figure*}
	\centering\includegraphics[width=12cm]
	{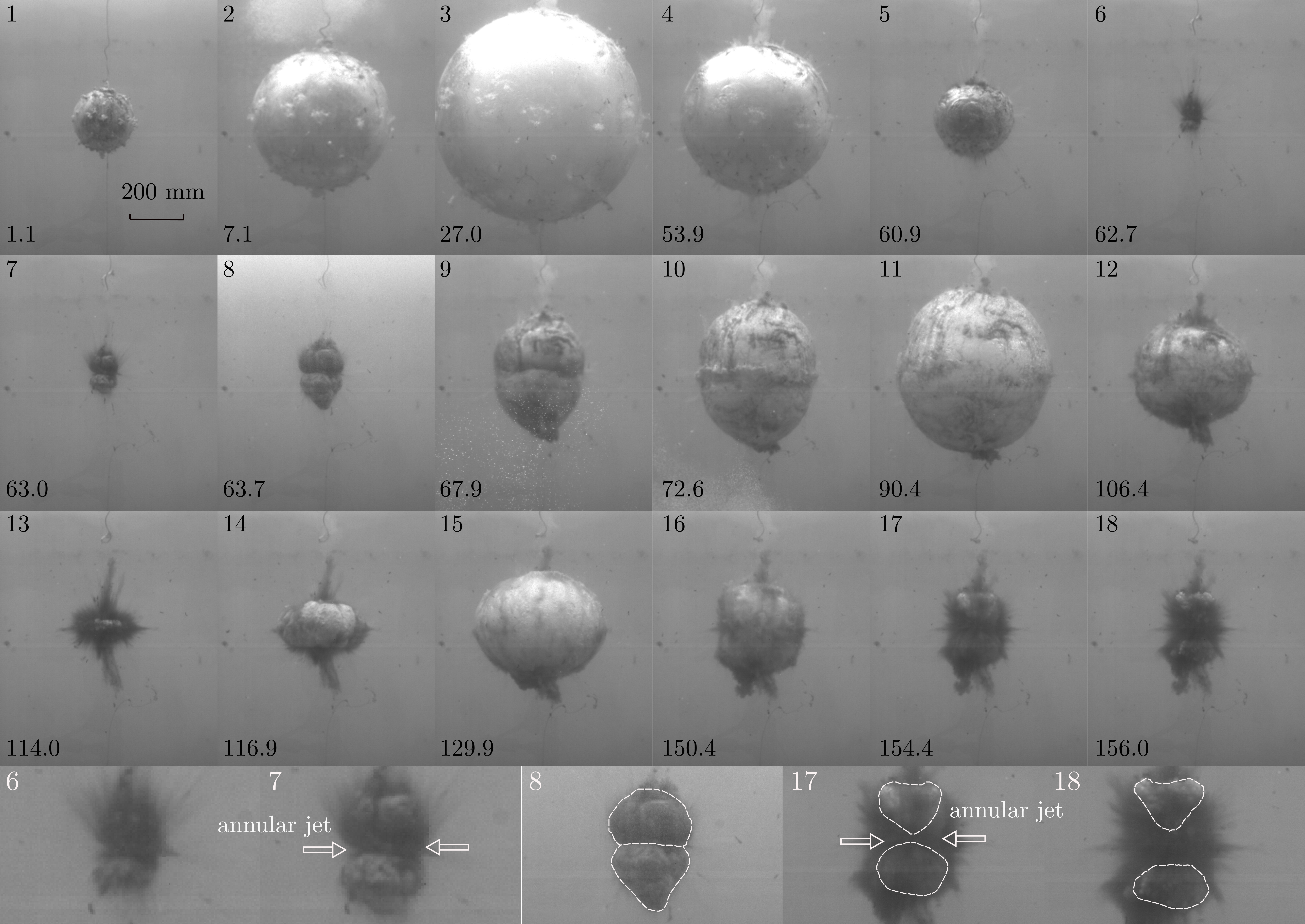}\\\caption{The bubble dynamics of the neutral collapse of  type $\rm\uppercase\expandafter{\romannumeral 1}$  within three bubble's oscillation. This bubble dynamics is characterized as an annular jet splitting the bubble at the end of the first and the second cycles (frame 7 and 17).}\label{Fig:free surface1.8-1}	
\end{figure*}
\begin{figure*}
	\centering\includegraphics[width=12cm]
	{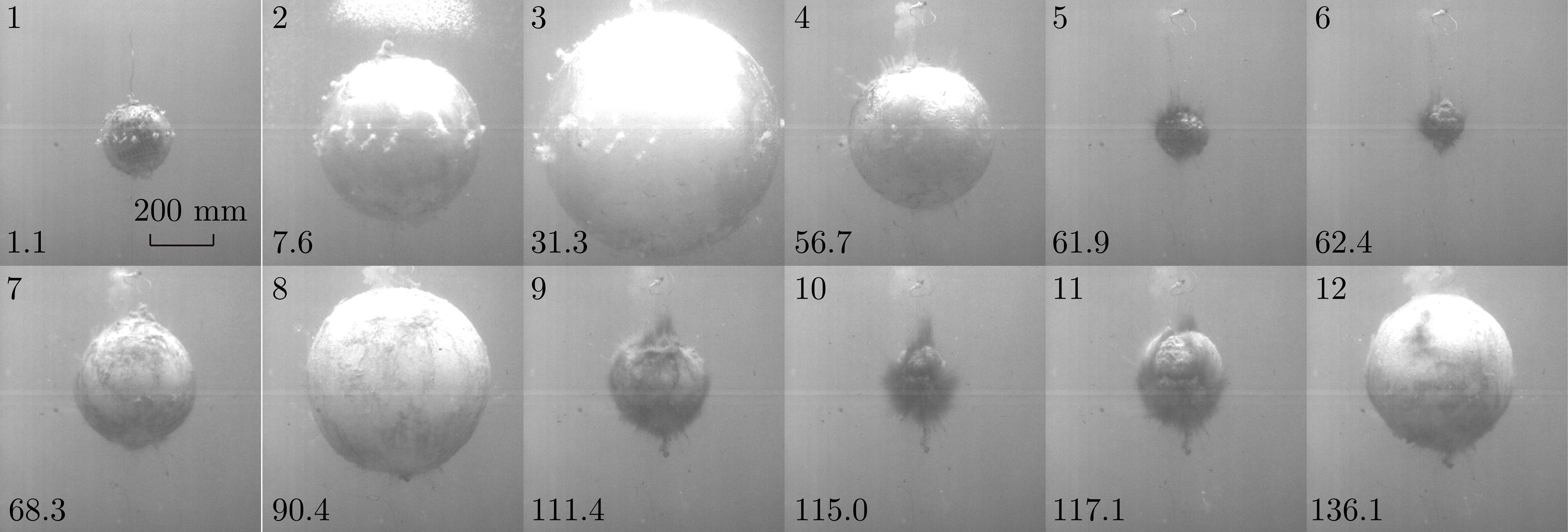}\\\caption{The bubble dynamics of neutral collapse of type $\rm\uppercase\expandafter{\romannumeral 2}$. The bubble nearly oscillates spherically and there is no obvious jetting phenomenon observed. The bubble slightly migrates upward at the effect of buoyancy.}\label{Fig:free surface1.8-2}	
\end{figure*}
\begin{figure}
	\centering\includegraphics[width=7cm]
	{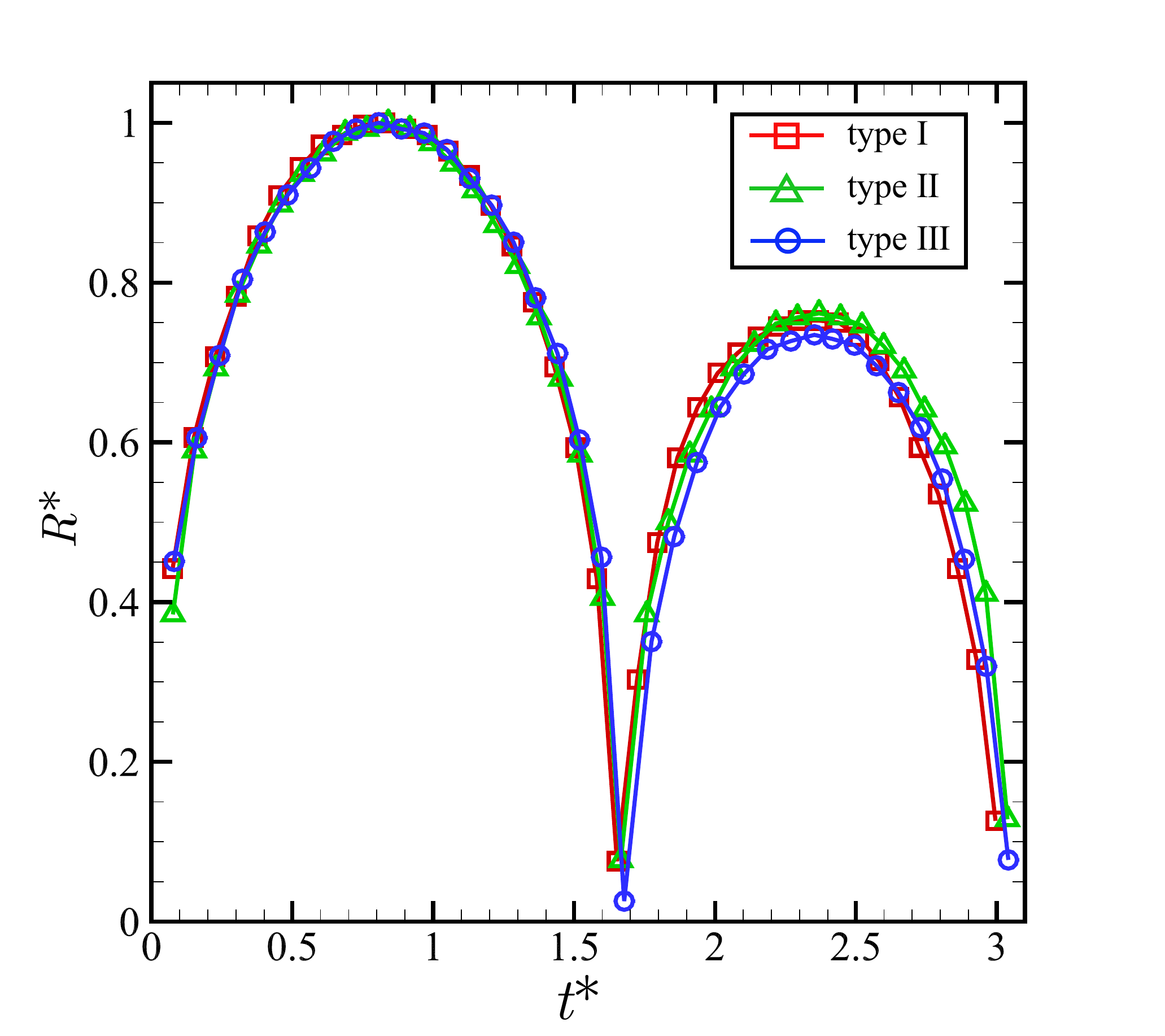}\\\caption{The time variation of the non-dimensional bubble radius in two bubble's oscillation. The three curves nearly overlap with each other in the first cycle and their differences are only revealed at the second cycle.}\label{Fig:r neutral collapse}	
\end{figure}
\subsection{Quasi-free field movement}
Once the standoff distance is larger than double the maximum equivalent radius of the bubble, such as $\gamma>2.1$, the bubble dynamics is similar to that in a free field. The discrepancy is that the migration of the bubble, as well as the upward migration speed at the end of the second cycle, is not as large as that in the free field ($H=2$m). On the one hand, though the bubble jets upward during the 1st cycle, it doesn't mean that the repellent force from the free surface is negligible: the repellent force is smaller than the buoyancy force. On the other hand, the bubble’s migration in the first cycle should be considered for its depth decreases as the bubble migrates. For the free surface cases, the bubble centroid migrates upward during the first cycle. This results in the  strengthened repellent force from the surface, which further influences the bubble's migration.

\section{More discussions}
\subsection{Migration of the bubble }

The migration of the bubble is an indicator of the magnitude of the combined effect of the Bjerknes effect and the buoyancy to the bubble. The migrations of the bubble centroid at different standoff parameters during the two consecutive cycles are shown in Fig.\ref{Fig:migration}. The bubble nearly assumes a spherical shape when it first expands, during which process the centroid nearly doesn't move except for the cases when $\gamma<$1.79 because of the upward direction of the overall pressure gradient along the axis. The direction and strength of the jet at the end of collapse determine the corresponding migration direction and the migration speed. At the rebounding stage of the second cycle, the migration speed decelerates. The relatively large discrepancies in the three repeated experiments at $\gamma=2.1$ again illustrate the instability of the bubble when contracting to the minimum at this depth.
\begin{figure*}
	\centering\includegraphics[width=12cm]
	{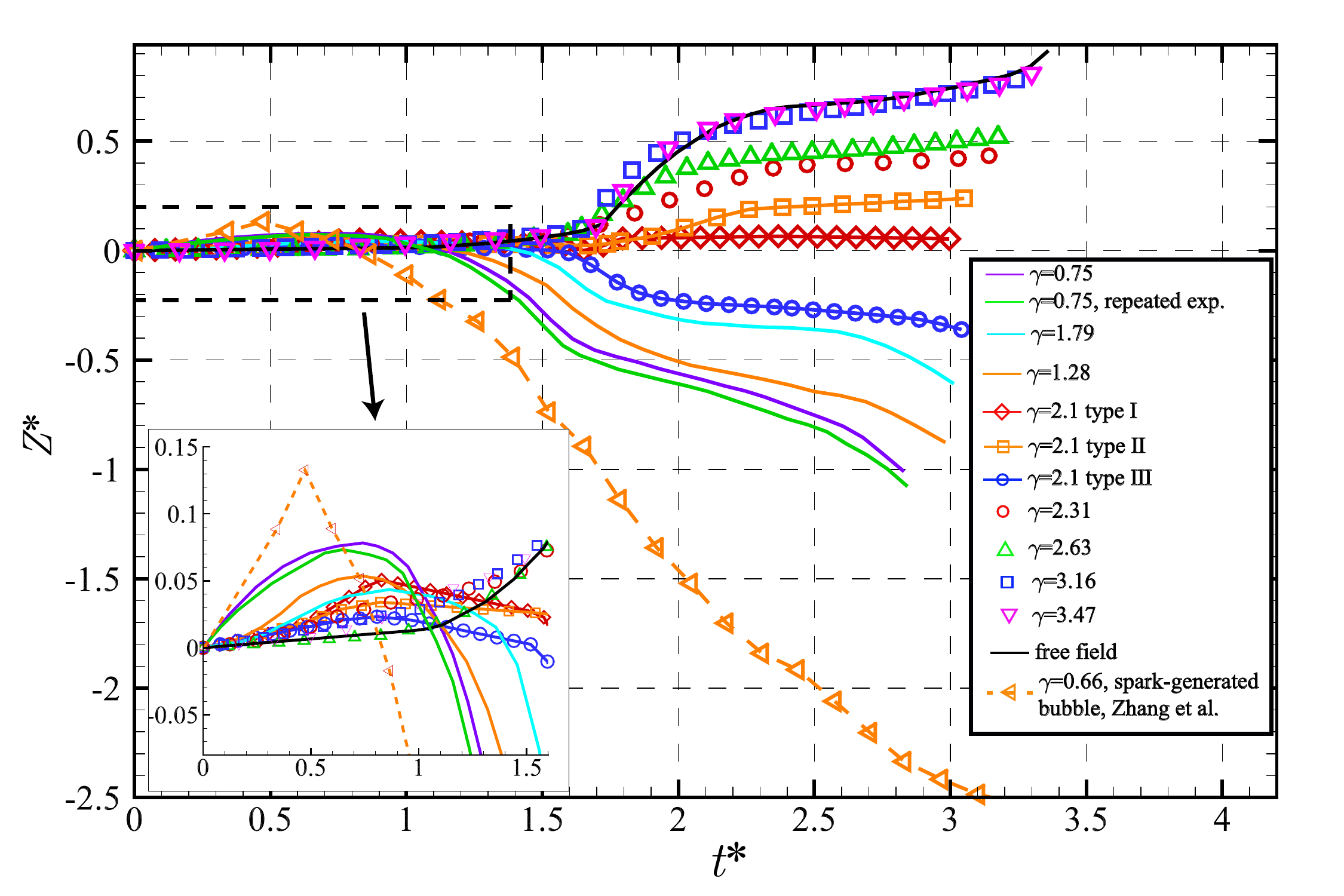}\\\caption{The migration of the bubble centroid at different standoff distance $\gamma$. The bubble in the free field as well as the one that moves downward ($\gamma<2.1$)  is presented with solid line. The upward movement ($\gamma>2.1$) is presented with the separated points and the bubble migrations at the transitional region ($\gamma\approx 2.1$) are presented with the solid line combined with the separated points. The spark-generated bubble experimental data from Zhang et al.\cite{RN2608} is added for comparison.}\label{Fig:migration}	
\end{figure*}

Generally, the free field experiments are conducted deep enough ($\gamma>7$) below the surface to get rid of the free surface effect whether it is an underwater explosion bubble, spark-generated bubble, or laser-induced bubble. However, there is no quantitative standard existed to determine the depth. Currently, there is no agreement on how to identify whether the bubble dynamics are influenced by the free surface or not. \citet{RN2735} postulated that the free surface is not necessary to be considered when the re-entrant jet is suppressed during the first collapse phase. Based on this criterion, $\gamma=2.1$ is a limit of the standoff distance after which the free surface effect is negligible as demonstrated in our experiment. But we estimated that this criterion is still a little robust. The re-entrant jet is suppressed means that the Bjerknes force and buoyancy have the same magnitude. However, it still indicates some involvement of the free surface effect. By comparing the time history of the migration curves of free surface cases with that of the free field experiment, we noticed that the bubble migration curve at $\gamma=3.16$ and $3.47$ match well with that in the free field. For quantitative analysis, we computed the time integral of displacement, i.e.: the area below the migration curve for $\gamma=3.16, \, 3.47$, and free field experiment. We found that the relative errors among them are within the limit of $3\%$. Also, it is explained in section.\ref{s:period} that the non-dimensional bubble oscillation periods of these two cases are close to that in the free field condition. The discrepancies among them are simply due to the difference in ambient hydrostatic pressure. This quantitative comparison concludes that $\gamma=3.16$ is the critical standoff distance at which  the free surface is not necessary to be considered. This standoff distance is mainly dependent on the specific condition, such as buoyancy effect, viscosity, etc. If the buoyancy parameter decreases, this critical standoff distance is supposed to increase as the buoyancy effect is weakened. In the experiments by Kannan et al.\cite{RN2735}, this critical standoff distance is obtained at $\gamma=6$ for deionized water based on their criterion that the re-entrant jet being suppressed in the first collapse phase is the condition for ignoring the free surface effect. The size of the maximum bubble radius in their experiment was a few millimeters, at which the buoyancy effect was considered to be significantly smaller than in our experiments. Their results are in accordance with our assumption. 

{ In Fig.\ref{Fig:migration}, the spark-generated bubble experiment at standoff distance $\gamma=0.66$ from Zhang et al.\cite{RN2608} is added for comparison. The maximum equivalent bubble radius in Zhang et al.\cite{RN2608} is about 27.75 mm, which is significantly smaller than the UNDEX experiments in our study ($\sim $0.4 m). By comparing the spark-generated bubble experiment at $\gamma=0.66$ with the UNDEX experiment at $\gamma=0.75$, it is found that the migration magnitude for the smaller scale bubble is relatively larger than the UNDEX bubble not only in the initial moving upward stage but also in the later consecutive downward migration stage. As the buoyancy effect is significantly smaller for the spark-generated bubble, the migration of the bubble is more susceptible to the free surface at nearly the same standoff distance $\gamma$.}
\subsection{Bubble's oscillation period}
\label{s:period}
The bubble's oscillation period is the quantity that researchers are particularly concerned about. Especially the relation between the bubble's oscillation period and the natural frequency of the structure due to the possibility of resonance. Some researchers have investigated the bubble's oscillation period when the bubble is initiated around the boundary \citep{RN2621,RN2603,RN2608}. Generally, the free surface will decrease the bubble's oscillation period, indicating faster bubble expansion and contraction.
{
Like the derivation process by Rayleigh\cite{RN2474} in the free field condition, a Rayleigh-like period\cite{RN3019} can be determined for the standoff distance $\gamma$ by taking the free surface condition into consideration, see Eq.\eqref{eq:Rayleigh-like period}. 
\begin{equation}
\label{eq:Rayleigh-like period}
 T^*=\sqrt{6}\int_{0}^{1}\sqrt{\frac{x^3}{1+ x^3}(1-\frac{x}{2\gamma})}\, dx
\end{equation}
 To our knowledge, unlike the Rayleigh period, the analytical solution can't be obtained for a random standoff distance $\gamma$ in Eq.\eqref{eq:Rayleigh-like period}, rather it is solved numerically. The fitting curve of the Rayleigh-like period for Eq.\eqref{eq:Rayleigh-like period} with at the range $1 \le\gamma\le 3.2$ is
\begin{equation}
\label{eq:fitting curve}
T^*=0.012\gamma^5-0.1479\gamma^4+0.73\gamma^3-1.85\gamma^2+2.51\gamma+0.151
\end{equation}
}
\begin{figure}
	\centering\includegraphics[width=7cm]
	{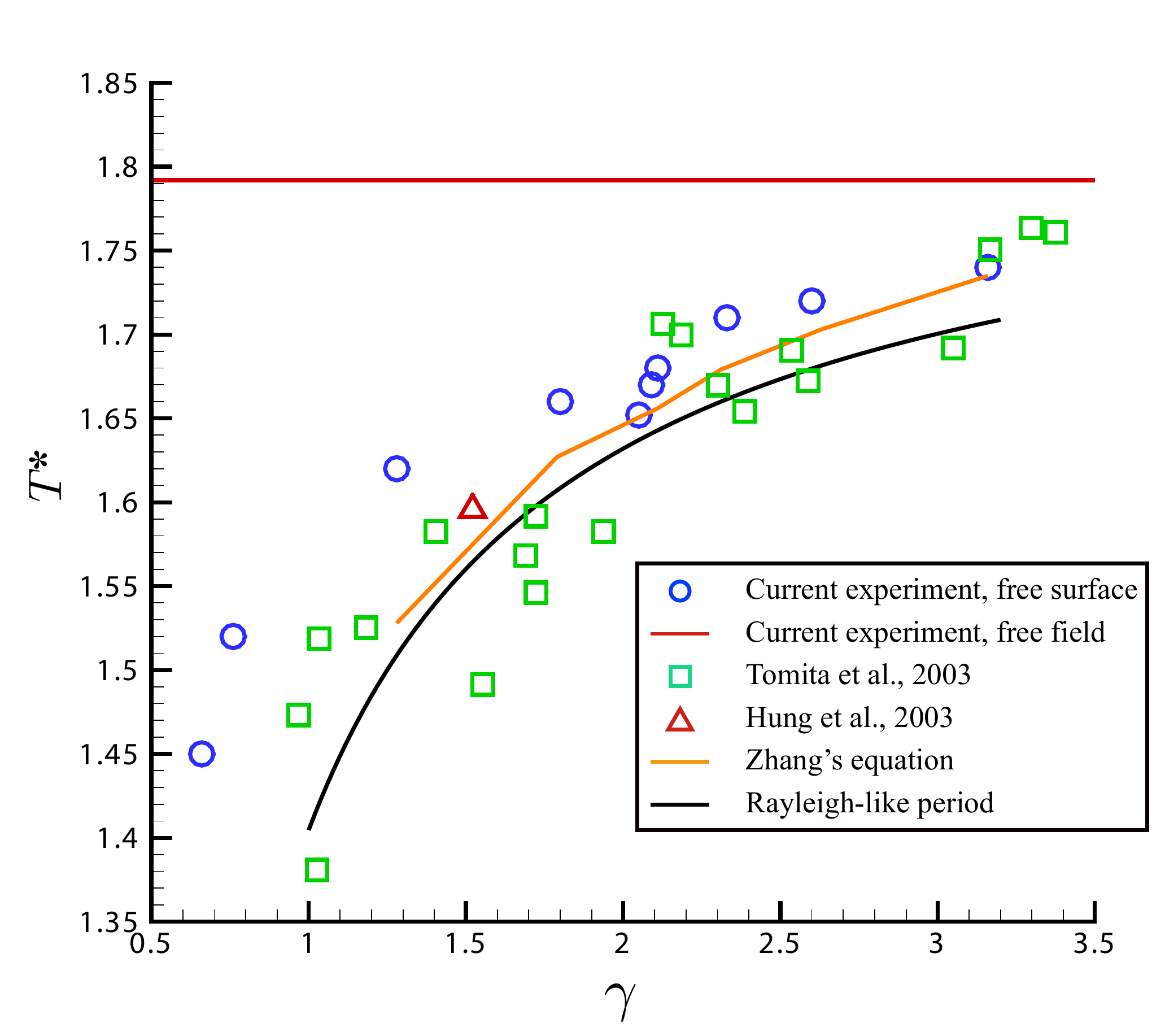}\\\caption{The first bubble oscillation period at different standoff distance $\gamma$. The free surface experiments and the free field experiment are all presented and compared with that of the laser-induced bubble experiment \citep{RN2614}, the mini-charge UNDEX experiment \citep{RN2603} as well as equation\eqref{eq:unified}.}\label{Fig:period}	
\end{figure}

Fig.\ref{Fig:period} shows the variation of the bubble's oscillation period with standoff distance $\gamma$ from experiments as well as from Eq.\eqref{eq:unified} and Eq.\eqref{eq:fitting curve}. The results from Eq.\eqref{eq:unified} are calculated considering the optimal parameter derived from the free field experiment: $C_d=1.5$ and $C_a=0.2$. It shows that the non-dimensional bubble's oscillation period increases with the standoff distance and the theoretical model (Eq.\ref{eq:unified}) can reliably predict the bubble's oscillation period. The Rayleigh-like period is a bit underestimated than the results from Eq.\eqref{eq:unified}. The reason for the underestimated results can be that the internal bubble pressure is not considered in the derivation process and it can be used as a reference period just like the Rayleigh period in the free field condition. Contrary to the assumption that the pressure remains constant in the infinity for the free field condition, the air remains at atmospheric pressure in the vicinity of the bubble when the bubble is near a free surface. This results in the increased external pressure around the bubble that makes the bubble contract more swiftly. We suppose this accounts for the decreased bubble's oscillation period when the detonation position is closer to the free surface. By comparing it with experimental data from other data in the existing literature, it shows that our free surface oscillation period is generally larger than the data from Ref.\cite{RN2614}, in which the bubble was induced by laser. It may be due to the dissolvability of the bubble's internal gaseous content. The bubble's internal content from spark-generated bubbles or laser-induced bubbles tends to disintegrate into the surrounding water under high internal pressure. This mechanism reduces the overall internal energy which in turn decreases its ability to resist external pressure. It should be considered that the non-dimensional oscillation period in the free field case (1.79) is slightly lower than the Rayleigh period (1.83). We postulate that it is caused by the reflected wave from the tank walls. As this discrepancy is not significantly large enough. Therefore, we focused only on the free surface effects which are much more influential on the bubble dynamics, thereby neglecting the boundary effect. 

\subsection{Pressure characteristics near a free surface}
The characteristics of pressure for shock waves and bubble pulses have always been of research interest because they are the direct loads on the floating structures. In the free field condition, the bubble is free from the influence of the boundary, and the pressure at one fixed point decreases exponentially with time for the shock wave, see the experimental results in Cui et al. (2016)\citep{RN33} and the empirical formulas in \citet{RN2836} (1973). If the free surface is taken into consideration, a rarefaction wave is reflected because of the significant difference in acoustic impedance between air and water. The rarefaction wave magnitude decreases the local pressure significantly below saturation pressure causing the bulk cavitation under the free surface, as shown in Fig.\ref{Fig:free surface0}. In this section, we are presenting the pressures measured near the free surface with two pressure sensors. The test conditions and positions of these two pressure sensors or the gauge points are shown in Table.\ref{table:pressure measurement}. 
\begin{table*}
	\centering
	\caption{The setups for the pressure sensors.}
	\begin{ruledtabular}
	\begin{tabular}{ccccccccc}
		\multirow{4}{*}{NO.}&\multirow{4}{*}{\makecell[c]{depth\\(m)}}&\multirow{4}{*}{\makecell[c]{standoff\\parameter}}&\multicolumn{3}{c}{sensor 1 (S1)}&\multicolumn{3}{c}{sensor 2 (S2)}\\
		\cline{4-9}
		&&&\multirow{3}{*}{\makecell[c]{depth (m)}}&\multirow{3}{*}{\makecell[c]{horizontal distance (m)}}&\multirow{3}{*}{\makecell[c]{distance (m)}}&\multirow{3}{*}{\makecell[c]{depth (m)}}&\multirow{3}{*}{\makecell[c]{horizontal distance (m)}}&\multirow{3}{*}{\makecell[c]{distance (m)}}\\
		&&&&&&&&\\
		\hline
		&&&&&&&&	\\
		1&0.3&0.75&0.4&0.75&0.76&0.35&0.55&0.55\\
		2&0.3&0.75&0.4&0.75&0.76&0.35&0.55&0.55\\
		3&0.5&1.28&0.67&0.8&0.82&0.52&1.3&1.3\\
		4&0.7&1.79&0.7&0.8&0.8&&&\\
		5&0.8&2.05&0.7&1.0&1.0&&&\\
		6&0.8&2.1&0.67&0.8&0.81&0.52&1.3&1.33\\
		7&0.8&2.1&0.35&0.55&0.71&0.4&0.75&0.85\\
		8&1.0&2.63&0.6&1.0&1.08&0.4&0.8&1.0\\
		9&1.2&3.16&0.25&1.0&1.38&0.5&0.6&0.92\\
		10&2.0&5.71&1.35&0.9&1.11&0.5&0.9&1.75\\
	\end{tabular}
\end{ruledtabular}
	\label{table:pressure measurement}
\end{table*}
Fig.\ref{Fig:pressure-1-0} shows the time history of the measured pressures at $\gamma=2.63$, in which the shock wave, first bubble-induced pulse, and second bubble induced pulse have been captured precisely. In Fig.\ref{Fig:pressure-1-0}(b), the graph showed an abrupt jump in pressure and this jump could be due to the arrival of the rarefaction wave. This can be verified theoretically. The reflected rarefaction wave can be deemed to be radiated from a fictitious charge by mirroring the real charge at the surface. The speed of a rarefaction wave can be approximated by the speed of sound in the water. By calculating the distances between two charges to the sensor, the time for the shock wave can be calculated analytically. The analytical value for the experimental setup of Fig.\ref{Fig:pressure-1-0} is estimated as 0.56 ms which is close to our experimental result of 0.62 ms. The second peak pressure after the cavitation is due to the reflected wave from the boundary as the time for its arrival matches well for a reflected wave from the tank walls. { For the initial stage of the shock wave (Fig.\ref{Fig:pressure-1-0}(c)), it shows that the pressure curve matches well with the empirical fitting curve \citep{RN2836}:
	\begin{equation}
	P=P_{\text{max}}e^{-t/\theta}
	\end{equation}
	 	in which $e$ and $\theta$ are natural exponential and the exponential damping constant, respectively. The exponential damping constant $\theta$ denotes the time that the shock wave pressure reduces from its maximum peak pressure $P_{\rm max}$ to the value of $P_{\rm max}/\rm e$, which reveals the shock wave damping characteristics at the early stage. Hence it is thought to be irrelevant to the standoff distance $\gamma$ and the bubble dynamics. Its value should be determined for the later time-integral calculation of pressure. Its value ranged from 0.02 ms to 0.036 ms in our experiments. The mean value of them was estimated as 0.028 ms, which was adopted during the later calculation. } For the bubble pulse pressure (Fig.\ref{Fig:pressure-1-0}(d)), a clear rising and falling trend of pressure graphs is observed for the duration of several milliseconds. While there exists a sudden rise and jump in the shock wave curve, and the lasting time for it is only half a millisecond (see Fig.\ref{Fig:pressure-1-0}(b)). Detailed information on these two types of pressure waves will be discussed separately. The decreased peak pressure between the 1st and the 2nd bubble pulse indicates that a portion of the bubble energy is lost during each collapse phase.
\begin{figure*}
	\centering\includegraphics[width=14cm]
	{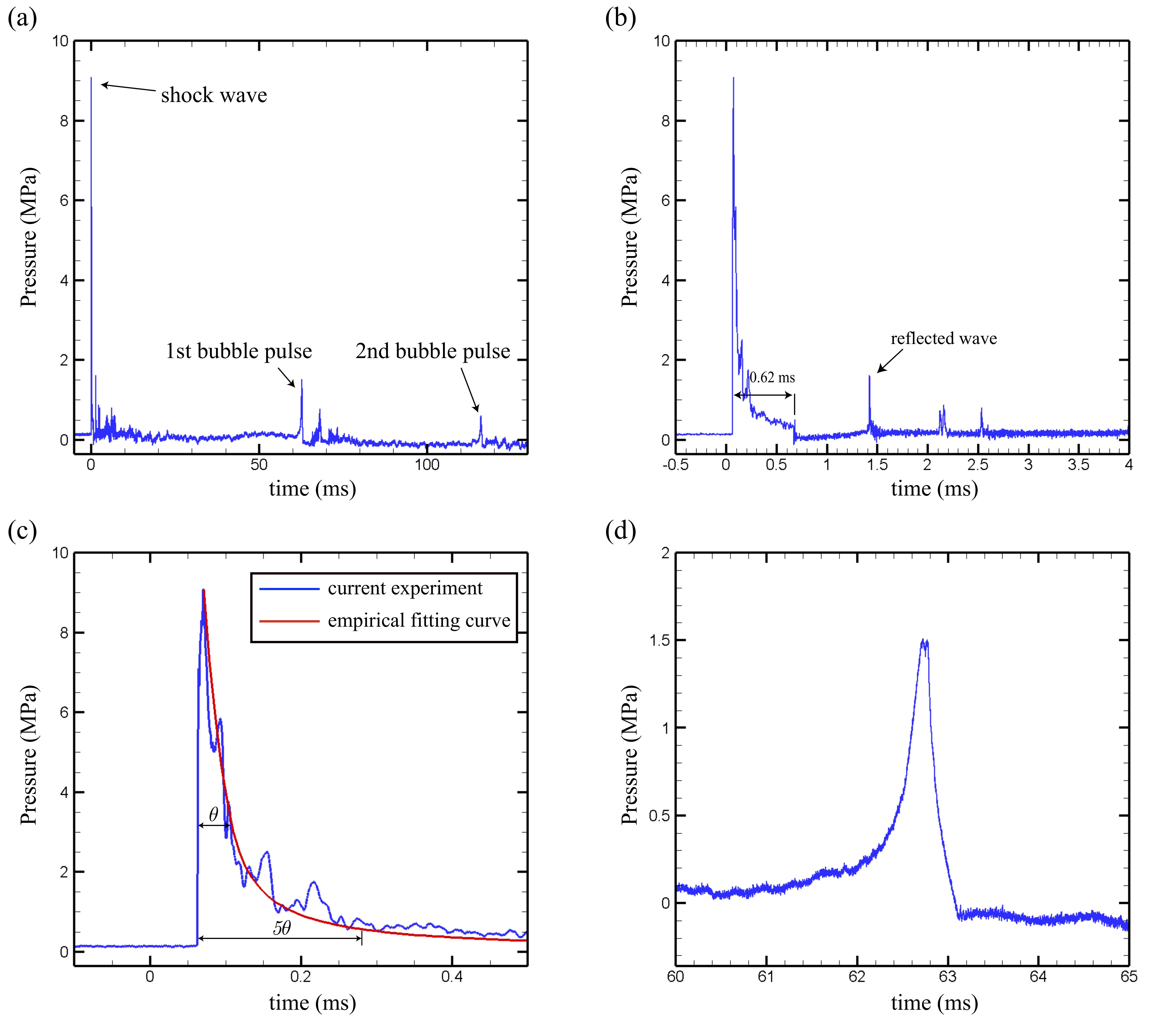}\\\caption{The measured pressure at the standoff parameter $\gamma=2.63$. The pressure sensor is located 1.08 m away from the charge. (a) The time history of the pressure wave in the whole process.  (b) The pressure wave in the shock wave process from the detonation. (c) The measured pressure and the pressure obtained from the empirical formula \citep{RN2836} in the early stage of the shock wave. (d) The time  history of the 1st bubble pulse induced by the bubble collapse.}	\label{Fig:pressure-1-0}
\end{figure*}

As it has been shown in Section.\ref{ss:burst}, we have observed two different phenomena at the depth $H=0.3$ m: droplets splashing on the bubble interface when the bubble collapses in one experiment and the absence of droplets splashing in the repeated experiment. The setups of pressure sensors for these two experiments were identical, which enables us to compare the influence of droplet splashing on the bubble's pulse pressure. The bubble pulse pressures measured by both sensors for these two experiments are compared in Fig.\ref{Fig:com0-3}. The comparison shows two kinds of totally different pressure curves. The time history of pressure at different distances is similar for both cases (see Fig.\ref{Fig:com0-3}(a)(c) and (b)(d)). For the case with droplets splashing on the bubble wall, the curve is oscillatory and contains multiple peaks. While for the case without droplets impacting on the bubble surface, the pressure graphs consist of a clear single peak. This single peak magnitude is larger than the former one at the same distance (see Fig.\ref{Fig:com0-3}(a)(b) and (c)(d)). By calculating the time integral of the pressure, i.e. pressure impulse, it is found that the impulse for the case without droplets impacting on the bubble wall is larger at the respective distance. We think that it might be caused by the disintegration of the bubble bulk into smaller daughter bubbles. Then, each bubble collapses separately accounting for the multi-peak in the curve.
\begin{figure*}
	\centering\includegraphics[width=12cm]
	{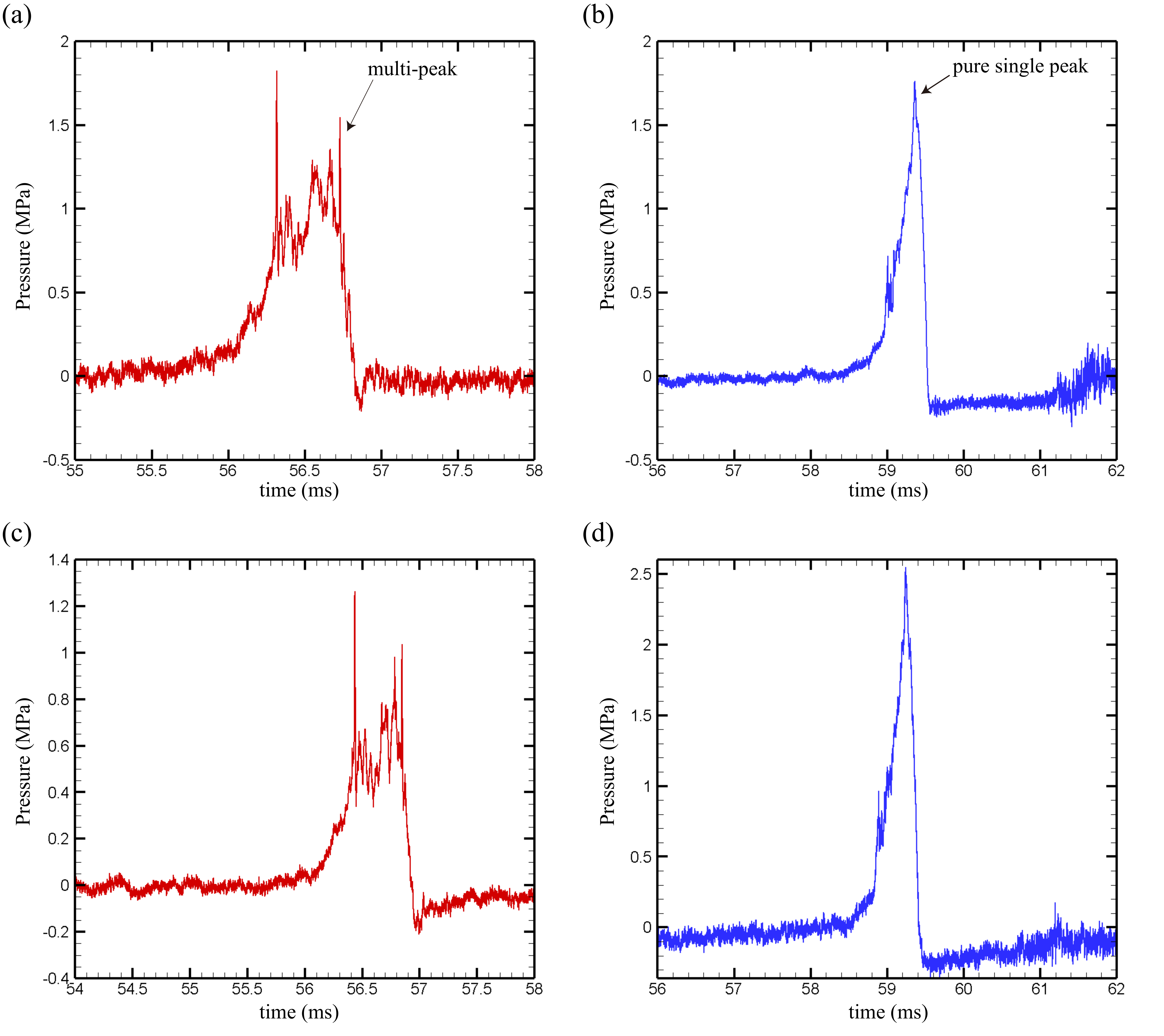}\\\caption{The measured bubble pulse pressure at the standoff parameter $\gamma=0.75$. (a)(c) The bubble pulse pressure at two sensor points S1 (radial distance 0.75 m) and S2 (radial distance 0.55 m) with droplets splashing on the bubble wall, respectively. (b)(d) The bubble pulse pressure at two sensor points S1 (radial distance 0.75 m) and S2 (radial distance 0.55 m) without droplets splashing on the bubble wall, respectively.}	\label{Fig:com0-3}
\end{figure*}

To comprehensively assess the action of pressure with time, the impulse is also taken into account, which is obtained via the following formula:
\begin{equation}
\label{eq:impulse}
I=\int_{t_{\text{lower}}}^{t_{\text{upper}}} P dt,
\end{equation}
where $t_{\text{lower}}$ and $t_{\text{upper}}$ denote the lower limit and upper limit of the integral range, respectively. As it is shown in Fig.\ref{Fig:pressure-1-0}(c), the shock wave pressure has subsided to a low level after a time interval of $5\theta$ and the area below the pressure curve during this time interval has occupied the majority of the pressure impulse. Hence, $t_{\text{upper}}$ equals $5\theta$ in our study, which is also suggested by \citet{RN2475} (1948). According to our observation, the reflected rarefaction wave arrives much later than $5\theta$. Therefore the shock wave impulse is assumed to be unaffected by the rarefaction wave. The variations in the magnitude of peak pressure and impulse of the shock wave with gauge distance $r$ are shown in Fig.\ref{Fig:shock wave}. It shows that the peak $P_{\rm max}$ and the distance $r$ follow $1/r^{1.11}$ dependency by the regression analysis, which is remarkably close to the empirical rule of $1/r^{1.13}$ relation. As for shock wave impulse $I_s$, it is found that $I_s$ and $r$ follow a $1/r^{1.29}$ relation.
\begin{figure*}
	\centering\includegraphics[width=12cm]
	{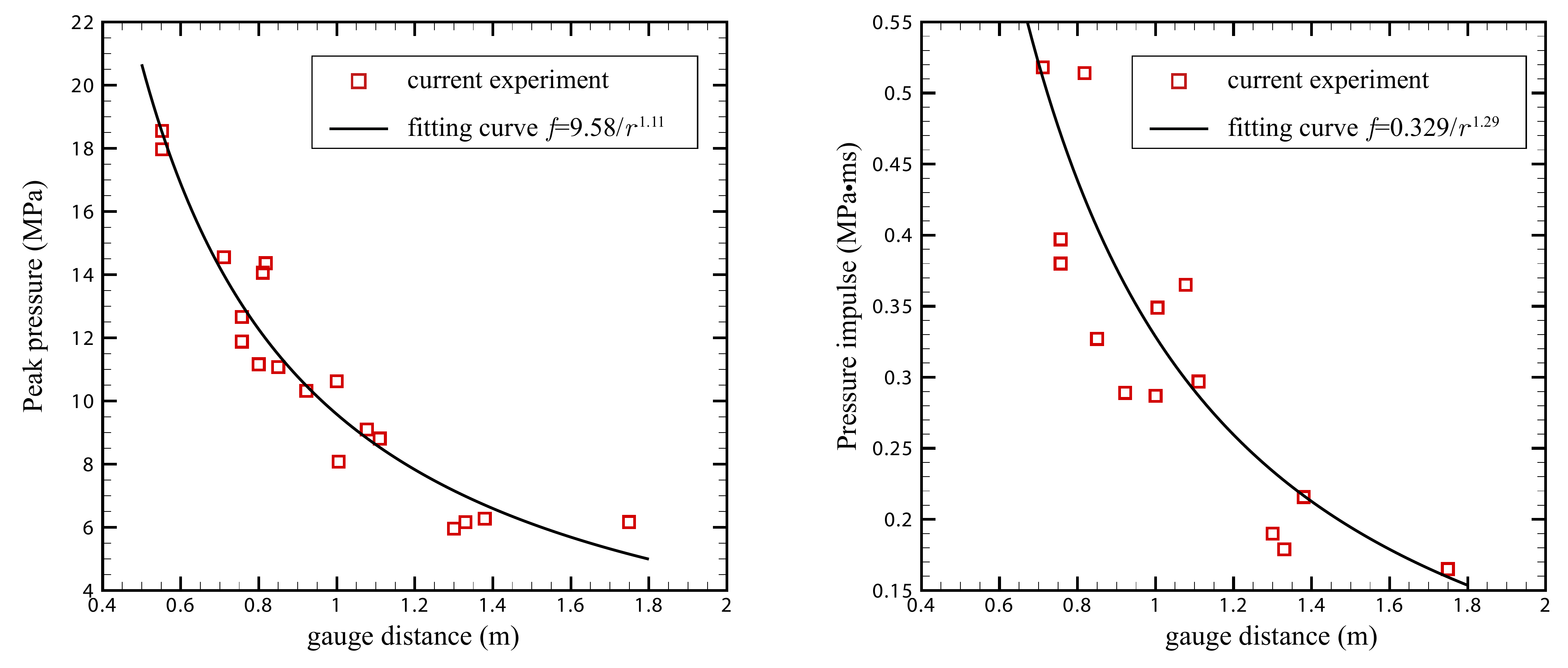}\\\caption{Right: shock wave peak at different gauge distances. Left: shock wave impulse at different gauge distances.}	\label{Fig:shock wave}
\end{figure*}

Like the shock wave, the pressure characteristics for the bubble pulse can also be analyzed by the pressure peak and the pressure impulse. The estimation of pressure peak is relatively simple while there are two points needed to be considered for the impulse $I_{\rm b}$: the time integral limit and baseline for the pressure calculation. There is no strict rule applicable for the above two points just like the shock wave. As it is shown in Fig.\ref{Fig:free field pressure}, the pressure signal curve remains below zero for most of the time. Theoretically, the bubble impulse should be calculated when the pressure is above the hydrostatic pressure. However, the recorded pressure curves were generally oscillatory, which makes it uneasy to identify the time integral range ($t_{\text{lower}}$ and $t_{\text{upper}}$ in Eq.\eqref{eq:impulse}) to calculate impulse. In our study, $t_{\text{lower}}$ is chosen to be the time when there is an obvious rise in the pressure curve, and $t_{\text{upper}}$ is the time when pressure subsides to nearly the same pressure value at $t_{\text{lower}}$. According to the analysis of Cole\cite{RN2475} (1948), the time that pressure remains positive takes up $22\%$ of the bubble's oscillation period. While in our experiments, the integral range is about 2 ms at most which is $3\%$ of the bubble's oscillation period.{ To compensate this discrepancy, a lower limit value  $I_{b}^{\text{lower}}$ and an upper limit value $I_{b}^{\text{upper}}$ for the impulse are obtained by choosing two baselines. For the lower limit value $I_{b}^{\text{lower}}$, the baseline for the pressure is chosen to be zero at the time $t_{\text{lower}}$. And for the upper limit value $I_{b}^{\text{upper}}$, the baseline at the time $t_{\text{lower}}$ is 0.1MPa. The selected baseline does not affect the bubble pulse peak but influences the time integral quantities. The ultimate impulse $I_b$ is the mean value of the  above-mentioned results, i.e. $(I_{b}^{\text{upper}}+I_{b}^{\text{lower}})/2$. }As the charge weight is identical in all our experiments, the reduced pressure peak $P_{\rm max} \cdot r$ and reduced impulse $I_{\rm b} \cdot r$ are used to investigate the relation of $P_{\rm max}$ and $I_{\rm b}$ with gauge distance $r$ at different standoff distance $\gamma$, which is shown in Fig.\ref{Fig:bubble pulse}.
\begin{figure*}
	\centering\includegraphics[width=14cm]
	{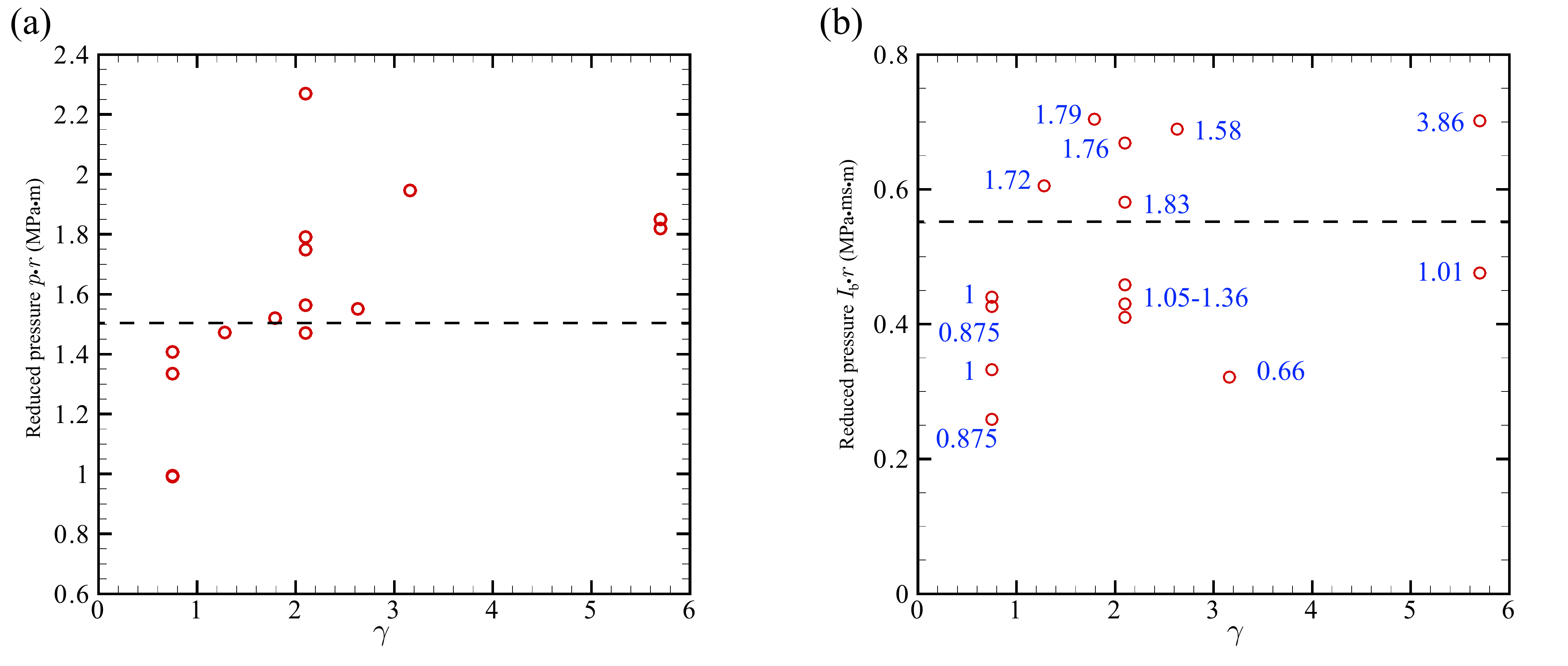}\\\caption{(a) The reduced bubble pulse peak pressure $P_{\rm max} \cdot r$ against the standoff distance $\gamma$. The dashed line denotes the mean value of the concentrated points. (b) The reduced bubble pulse impulse $I_{\rm b} \cdot r$ against the standoff distance $\gamma$ with non-dimensional gauge depth marked aside. The dashed line divides the experimental data into two parts: one part is the region where the bubble pulse impulse has been reduced by the rarefaction wave from the free surface; the other is region where the calculated bubble pulses impulse are not affected by the rarefaction wave.}	\label{Fig:bubble pulse}
\end{figure*}
As it can be seen in Fig.\ref{Fig:bubble pulse}(a), the experimental data for reduced pressure are mainly distributed along the dashed line $P\cdot r=1.5$, which is the calculated mean value of these concentrated points. This indicates a roughly $1/r$ relation between $P_{\rm max}$ and $r$. But it can also be seen that the points at $\gamma=2.1$ are higher than the dashed line. Referring to the aforementioned bubble dynamics patterns, it can be seen that the bubble collapses neutrally at $\gamma=2.1$ where Bjerknes force and buoyancy are roughly balanced. As has been mentioned by Brett et al.\cite{RN2631}, a local high-pressure is also captured at the point where the bubble nearly remains stationary. The discrepancy is that this point in Brett et al.\citep{RN2631} is not related to the neutral collapse point as referred to in our experiments. It has been observed that the highest point at $\gamma=2.1$ is from the case that the bubble collapses spherically(typeII of the neutral collapse, see Fig.\ref{Fig:free surface1.8-2}). This feature can be attributed to the full compression of gaseous contents inside the bubble \citep{RN2631} which can partially support the energy loss mechanism. This energy loss phenomenon is discussed in section\ref{s:energy loss}.  

For the reduced impulse shown in Fig.\ref{Fig:bubble pulse}(b),  there is no line that most experimental data reached. It can be observed that the higher pressure peak doesn't guarantee a higher impulse. The obtained experimental data can be divided into two regimes by the dashed line in Fig.\ref{Fig:bubble pulse}(b):  the points above the dashed line can be considered to be distributed along a line between them, which means that these points follow the $1/r$ relation derived in Cole\cite{RN2475} (1947). While the points below the dashed line are scattered independently. With the non-dimensional gauge depth scaled by the maximum radius marked aside, it can be noticed that the gauge depth may be responsible for the points distribution stated above. For example, the gauge points above the dashed line are all placed deep enough while the gauge points below the dashed line are all placed closer to the free surface. The most contrasting examples are the case in the free field ($\gamma=5.7$). The bubble pulse peak for the two gauge points follows the $1/r$ relation very well, while the impulse of the smaller gauge depth (1.01) is significantly smaller than that of the larger gauge depth (3.08). It shows that the free surface has an enormous influence on the bubble impulse pressure magnitude. As the recorded wave profile is the superposition of the direct wave emitted by the bubble and the reflected wave from the boundary, 
the bubble impulse pressure magnitude is thought to be mainly influenced by the rarefaction wave from the surface. Unlike shock waves, the bubble pulse has a much wider pulse width, which makes it vulnerable to the reflected rarefaction wave. When the gauge point is close to the free surface, the reflected rarefaction wave follows the direct wave right after. From Fig.\ref{Fig:bubble pulse}, we can see that the non-dimensional depth $h^*=1.58$ is the critical depth that the reflected wave does not influence the magnitude of impulse in our experiments. When the gauge point is deeper than this depth, the bubble impulse $I_b$ with gauge distance $r$ again conforms to the $1/r$ relation.

The ratio of the bubble impulse and the shock wave impulse ranges from 0.84 to 2.1 (most of the data are above 1). It seems that the impulse magnitude for the bubble pressure pulse is generally larger than that of the shock wave magnitude. The mechanisms of impulse and shock wave emission are of research interest and need to be taken into consideration during the underwater explosion process to comprehensively analyze the loads.
\subsection{Energy loss during the first collapse }
\label{s:energy loss}
It has been indicated by Lee et al.\cite{RN2622}, most of the energy of the bubble  is radiated out in the form of a pressure wave, which may cause severe damage to the nearby structure. It has been observed in our experiments that the second bubble pulse is either too small or is not being measured by pressure sensors. It indicates that the remaining energy of the bubble during the second oscillation cycle is nearly negligible compared with the first collapse. This reveals that the majority of energy loss takes place during the first collapse, which is the focus of the current study. 

Here, the energy loss parameter is defined as $\alpha=1-E_{n+1}/E_{n}$, in which $n$ denotes the oscillation cycle number. Then the energy loss due to bubble collapse can be calculated via the following two formulas.
\begin{equation}
\label{eq:energy loss volume}
\alpha=1-\frac{V_{n+1}}{V_{n}}
\end{equation}
\begin{equation}
\label{eq:energy loss period}
\alpha=1-(\frac{T_{n+1}}{T_{n}})^3
\end{equation}
Fig.\ref{Fig:energy_loss} presents the calculated results of the energy loss parameter during the first collapse against the standoff distance based on both, the volume-based and the period-based approaches. It shows that though the absolute values for two approaches at the respective standoff distance $\gamma$ are different, the overall tendencies of $\alpha$ against $\gamma$ are fairly identical: Energy loss parameter $\alpha$ increases with an increase in $\gamma$ until the neutral collapse limit point $\gamma=2.1$. After that, it decreases with further increases in $\gamma$.
\begin{figure}
	\centering\includegraphics[width=7cm]
	{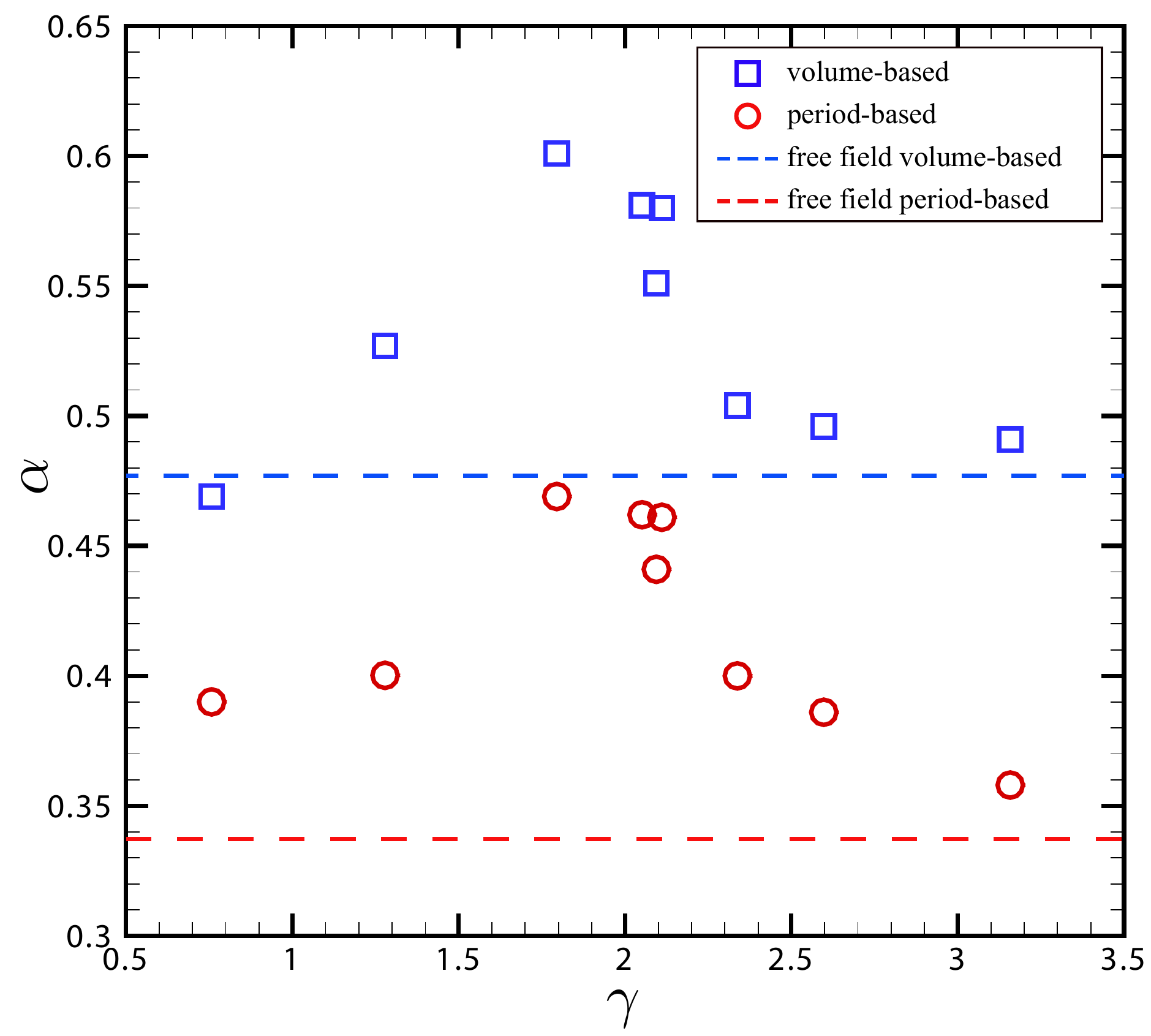}\\\caption{The variation of the energy loss paramiter during the first collapse with the standoff distance $\gamma$ based on the period-based approach and the volume-based approach, respectively. The dashed lines denote the energy loss parameter in the free field. The tendencies based on the two approaches are the same though there are some discrepancies on the specific values.}\label{Fig:energy_loss}	
\end{figure}

A reasonable way to explain this phenomenon is that the free surface influences energy loss by affecting the bubble dynamics. 
In the process of bubble contracting to the minimum volume, i.e. collapse, $E_p^g$ is released to be transformed to the potential energy of the bubble $E_p^b$ and kinetic energy of the liquid $E_k$. As the bubble over-contracts due to inertia, part of $E_k$ turns to $E_p^b$. When the bubble starts to rebound, the internal pressure of the bubble is comparatively higher than the ambient pressure of the liquid. Such extreme discontinuity generates the bubble's pulsation wave, and it takes up some portion of energy loss which is derived from $E_p^b$. The more kinetic energy is transformed into the internal energy of the bubble, the more energy tends to be dissipated by the pulse wave. Accordingly, we can see that when the bubble is close to the surface, a strong jet is supposed to be produced which carries much kinetic energy. The migration curve in Fig.\ref{Fig:migration} can also reveal the kinetics of migration. As the detonation point goes deeper, the migration kinetic energy decreases and the internal energy increases correspondingly. At the point of neutral collapse, the vertical migration of the bubble is not large as is shown in Fig.\ref{Fig:migration}. The gaseous content inside the bubble can be fully compressed for a spherically oscillating bubble. That is the reason that the pressure sensor recorded the highest pulse pressure for the spherical oscillation bubble. After the point of neutral collapse, the free surface effect is reduced and the buoyancy starts to become dominant. The bubble forms the upward jet which in turn increases $E_k$ and decreases in $E_p^b$. It signifies that the reduced energy loss occurs when the detonation depth further increases. It should be noticed that more energy is lost at the depth $\gamma=1.79$, which is resulted from the jet carrying some portion of gas with it when the jet penetrates the bubble surface, see Fig.\ref{Fig:free surface1.6} (frames 7-8). It means that the loss of mass or explosive content is also an important source of energy loss.

 Here the proportion of energy carried by the shock wave is assessed by Eq.\eqref{eq:cal energy}. As the recorded pressure is the superposition of the direct wave and the reflected wave, $E_w$ will be overestimated for the wave reflected from the rigid boundary or underestimated for the rarefaction wave from the surface. As has been analyzed before, we think that the points above the dashed line in Fig.\ref{Fig:bubble pulse}(b) are unaffected by the free surface, which are selected to calculate $E_w$. The ultimate calculated results are shown in Table.\ref{table:energy ratio}.
\begin{table}
	\centering
	\caption{The calculated energy loss based on the volume-ratio approach at several depths and the calculated energy loss based on the radiated pressure wave.}
	\begin{ruledtabular}
	\begin{tabular}{ccccc}	
		Depth $H$ (m)& $\gamma$& $\Delta E$ (J) & $E_w$ (J)& $\frac{E_w}{\Delta E}$\\
		\hline
		0.5& 1.28&12644&6473-7927&0.512-0.627\\
		0.7& 1.79&15082&6785-8214 &0.450-0.545\\
		0.8& 2.1&14505&7557-9442&0.521-0.651\\
		0.8& 2.1&12994&8927-9954&0.687-0.766\\
		1& 2.63&11637&7576-8623&0.651-0.741\\
		2& 5.71&8634&5992-7012&0.694-0.812\\
	\end{tabular}
	\end{ruledtabular}
	\label{table:energy ratio}
\end{table}
\noindent $\Delta E$ is the total energy loss at the 1st collapse based on the volume-ratio measurement and the values for $E_w$ include the upper limit and lower limit as are done in the calculation of impulse. The ratio $E_w/\Delta E$ indicates the proportion of the radiated energy of the pressure wave to the total lost energy. It shows that this ratio reveals some relevance to the depth, which resembles  the energy loss parameter against depth. The small portion of radiated energy at $H=0.7$ m may be due to the larger denominator $\Delta E$ which is caused by the loss of explosive product. If we look closer at the specific values, we can find that the difference between the upper limit and the lower limit of the ratio can reach as large as $10\%$, while all this difference only comes from the adoption of pressure baseline to calculate the integral in Eq.\eqref{eq:cal energy}. And it also shows that the radiated energy occupies approximately 60$\%$ to 80$\%$ of the total lost energy for the first collapse.

\section{Conclusion}
\label{s:conclusion}
A series of UNDEX experiments were conducted with varying standoff distance $\gamma$. The variations of the bubble collapse patterns, oscillation period, centroid migration, and the energy loss against the dimensionless standoff parameter $\gamma$ are systematically investigated. Additionally, the characteristics of the shock wave and bubble pulsation pressure beneath the free surface are also measured. The significant findings and conclusions drawn from our study are as follows:        

(1) Four patterns of UNDEX bubble dynamics are identified for different regimes of the standoff parameters: (i) $0 \le\gamma\le 0.41$: bubble bursting at the free surface,  (ii) $0.41<\gamma< 2.05$: bubble jetting downward, (iii) $2.05\le\gamma\le2.11$: neutral collapse, and (iv) $2.11<\gamma< 3.16$: quasi-free field movement of bubble. In our UNDEX experimental setup, $\gamma=3.16$ is thought to be the critical standoff distance limit at which the effect of the free surface on bubble dynamics is negligible, in terms of jet direction and centroid migration. 

(2) The oscillation period decreases with a decreasing detonation depth $\gamma$. A satisfactory agreement is obtained from Zhang equation\cite{RN3019} and our experimental data. Derived from Zhang equation\cite{RN3019}, the Rayleigh-like period can reliably predict the bubble oscillation period when the bubble is close to the free surface ($\gamma>1$).

(3) The decrease of the bubble pulsation pressure versus the distance $r$ follows a $1/r$ law except for the neutral collapse condition. The strength of the bubble pulse can be weakened by the disintegration of the integrated bubble into daughter bubbles. Additionally, the ratio of impulse for the bubble pulse and the shock wave is found to be between 0.84 and 2.1 (most data are above 1), which shows the importance of the bubble pulse in underwater explosions.

(4) The energy loss parameter $\alpha$ (defined as $\alpha=1-E^{2}/E^{1}$, where $E^{i}$ denotes the bubble energy during the $i$th cycle) increases with $\gamma$ until the neutral collapse position at $\gamma\approx2.1$, after which it decreases with $\gamma$. The loss of the explosive product is found to be an important source of the lost energy. Additionally, the proportion of the radiated energy to the total energy loss is found to increase with $\gamma$ while it reaches a local maximum value for the neutral collapse position. This proportion reaches 70$\%$ to 80$\%$ in the free field experiment. 

\section{Acknowledgement}
Special thanks should be given to  Dr. Liu Nian-Nian for his assistance to the experiments.
\section{Declaration of Interests}
The authors report no conflict of interest.

\nocite{*}
\bibliography{1}

\end{document}